# Computational Associative Memory with Amorphous InGaZnO Channel 3D NAND-Compatible FG Transistors

Chen Sun[1], Chao Li[2], Subhranu Samanta[1], Kaizhen Han[1], Zijie Zheng[1], Jishen Zhang[1], Qiwen Kong[1], Haiwen Xu[1], Zuopu Zhou, Yue Chen[1], Cheng Zhuo[2], Kai Ni[3], Xunzhao Yin[2#], and Xiao Gong[1]*

[1]National University of Singapore, Singapore.
[2]Zhejiang University, Hangzhou, Zhejiang, China.
[3]Rochester Institute of Technology, Rochester, NY, USA
[#]Email address: xzyin1@zju.edu.cn
[*]Email address: elegong@nus.edu.sg

## Abstract

3D NAND enables continuous NAND density and cost scaling beyond conventional 2D NAND. However, its poly-Si channel suffers from low mobility, large device variations, and instability caused by grain boundaries. Here, we overcome these drawbacks by introducing an amorphous indium-gallium-zinc-oxide (*a*-IGZO) channel, which has the advantages of ultra-low OFF current, back-end-of-line compatibility, higher mobility and better uniformity than poly-Si, and free of grain boundaries due to the amorphous nature. Ultra-scaled floating-gate (FG) transistors with a channel length of 60 nm are reported, achieving the highest ON current of 127 $\mu$A/$\mu$m among all reported *a*-IGZO-based flash devices for high-density, low-power, and high-performance 3D NAND applications. Furthermore, a non-volatile and area-efficient ternary content-addressable memory (TCAM) with only two *a*-IGZO FG transistors is experimentally demonstrated. Array-level simulations using experimentally calibrated models show that this design achieves at least 240× array-size scalability and 2.7-fold reduction in search energy than 16T-CMOS, 2T2R, and 2FeFET TCAMs.

## Introduction

Flash memories, which utilize the electron storage mechanisms either on a floating-gate (FG) or charge-trap (CT) layer, have been widely used in massive data storage due to the cost-effectiveness[1–4]. They also show great potential for many emerging applications, such as in-memory edge computing[5,6]. Vertically stacked NAND flash memories, or 3D NAND[7–9], has replaced planar NAND flash memories by overcoming many issues faced by 2D NAND when scaling beyond 20 nm technology node, including the requirement of the advanced lithography technique, increased cell to cell interference due to proximity effect, and more severe threshold voltage ($V_{TH}$) shift per electron injection caused by the random telegraphic noise[10–12]. 3D NAND provides a scalable path for continual density improvement by stacking more and more layers for higher memory density while also increasing the number of bits stored in a single cell. Highly promising as it is, 3D NAND also faces challenges during the z-direction scaling (i.e., stacking more layers), especially with the poly-Si channel. Firstly, the mobility of the poly-Si in the flash memories is typically less than 10 $cm^2/V \cdot s$ due to the disordered structure[13], resulting in a low cell current and insufficient sense margin, especially after stacking hundreds of layers[14,15]. Secondly, the existence of grain boundaries in poly-Si degrades $V_{TH}$ distribution and causes time-dependent $V_{TH}$ instability because of charge trapping at the grain boundries[16–18]. Thirdly, the grain boundaries would also introduce large device-to-device variation in electrical characteristics at ultra-scaled channel length ($L_{CH}$), constraining the number of layers stacked due to extremely challenging etching of high-aspect-ratio holes for the NAND string. Therefore, for sustainable 3D NAND scaling, alternative channel materials that can solve these challenges while being compatible with existing processes are highly desired.

Amorphous indium-gallium-zinc-oxide (*a*-IGZO), a metal-oxide semiconductor[19–21], has emerged as one of the promising channel alternatives to poly-Si in 3D NAND owing to its high mobility, free of grain boundaries due to the amorphous nature, wafer-size thin film formation with high uniformity[22,23]. Thin-film transistors (TFTs) with *a*-IGZO as the channel have been applied in displays, sensors, logic devices, and neuromorphic computing[24–29]. Moreover, *a*-IGZO that can be processed within the back-end-of-line (BEOL) thermal budget could enable ultra-high density monolithic 3D integration by vertically stacking multiple tiers of functional layers of devices and circuits[30–33]. In addition, our previous study for ultra-scaled *a*-IGZO TFT discovered that there was no noticeable degradation of mobility as the thickness of *a*-IGZO reduces from 6 to 3.6 nm[34], which is different from Si suffering from the dramatic reduction in mobility when the thickness is reduced to less than 5 nm[35]. By suppressing the short channel effects (SCEs), a thinner channel could enable more aggressive channel length scaling to even less than 10 nm[36,37]. Therefore, there is tremendous potential for realizing high-performance and ultra-scaled flash memories (FG or CT transistors) based on an *a*-IGZO channel. Although there have been several reports on *a*-IGZO-based flash transistors[38–40], the performance of the devices needs significant improvement, especially for lowering the operating voltage and scaling the channel length to tens of nanometers for high-density and high-performance 3D NAND applications.

In addition to its application as NAND memory storage, FG transistors also find emerging applications for in-memory computing with ternary content addressable memory (TCAM) as an important example[41,42]. TCAMs, as associative memories, can perform pattern matching in a massive parallel manner, which is highly desirable for IP packets forwarding in networking routers, database engines, as well as recently investigated compute-in-memory applications[42–44].

Various memory technologies can be harnessed for the TCAM design. The conventional TCAM cell is based on complementary metal-oxide-semiconductor (CMOS) technology and consists of 16 transistors (16T-CMOS), suffering from low area efficiency and high dynamical leakage power due to its volatility. Research studies employing emerging non-volatile memories (NVMs) for TCAM design have also been presented, such as resistive random access memory (ReRAM)[45,46] and ferroelectric field-effect-transistors (FeFETs)[47,48]. However, for these NVM-based TCAMs, a high thermal budget due to the high process temperature of Si technology and complex peripheral circuits to work with the low ON/OFF current ratio ($I_{ON}/I_{OFF}$) are necessary. Moreover, the large OFF current ($I_{OFF}$) of emerging NVMs limits the scalability of the NVM-based TCAM arrays. The scalability of a TCAM array is predominately determined by the $I_{OFF}$ of the transistors. As the TCAM word size increases, i.e., more transistors are connected in parallel on the same match line, the total $I_{OFF}$ in the TCAM word increases. This could discharge the match line even at a matching case, leading to a reduction of sense margin to distinguish between the match and mismatch scenarios and therefore causing possible function failure. The $I_{OFF}$ of transistors depends strongly on the thermionic leakage current and the tunneling current at the channel/drain junction[49]. Both current components have a negative exponential relationship with the energy barrier height $\Phi$, while the tunneling current is additionally influenced by the tunneling barrier width[50,51]. The wide bandgap (3.2 eV) of *a*-IGZO leads to a much greater built-in potential and a larger tunneling barrier height and width than Si, resulting in an ultra-low $I_{OFF}$[52]. Thus, TCAM based on FG transistors with an *a*-IGZO channel can overcome the aforementioned disadvantages of 16T-CMOS and NVM-based TCAMs. Firstly, it has a lower thermal budget and simplified peripheral circuits due to much larger $I_{ON}/I_{OFF}$ as compared with Si counterparts. Secondly, the ultra-low $I_{OFF}$ of *a*-IGZO transistors and 2T configuration in an *a*-

IGZO FG transistor-based TCAM cell could achieve much lower search energy than 16T-CMOS and NVM-based TCAM designs. Thirdly, with the extremely low $I_{OFF}$ of *a*-IGZO FG transistors, a much larger scale of the TCAM array can be realized, enabling applications that require a large array size, such as hyperdimensional (HD) computing[53,54].

In this work, we report the realization of *a*-IGZO FG transistors with outstanding electrical characteristics, including a low $I_{OFF}$ of less than $10^{-7}$ $\mu$A/$\mu$m, a large $I_{ON}/I_{OFF}$ of $1\times10^{8}$, a low subthreshold swing (*SS*) of ~80 mV/decade, and a high mobility of 32.6 $cm^2/V\cdot s$. Our *a*-IGZO FG transistors show a large memory window (MW) of 1.8 V at a low operating voltage with a retention time of more than 10 years, indicating the floating-gate-engineered nonvolatility. The endurance measurement of 1000 cycles of switching slightly reduces the MW from 1.8 to 1.6 V. From the evaluation of temperature-dependent operations, *a*-IGZO FG transistors still offer an MW of 1.4 V even at a high temperature of 80 °C. By scaling the $L_{CH}$ to 60 nm, record-high ON current ($I_{ON}$) of 127 $\mu$A/$\mu$m was realized and is much higher than reported flash transistors using an *a*-IGZO channel, while maintaining the MW larger than 1.5 V. Furthermore, an area-efficient TCAM cell with only two parallel-connected *a*-IGZO FG transistors was also demonstrated using a low thermal budget of 300 °C and functionalities of writing and searching are realized. Using the experimentally calibrated device parameters, we further performed array-level simulations to benchmark the scalability, search energy, and search delay of our *a*-IGZO FG transistor-based TCAM array with those TCAM arrays based on 16T-CMOS cell, two-transistor-two-ReRAM (2T2R), and 2FeFET. The results highlight that, compared with 16T-CMOS, 2T2R, and 2FeFET TCAM technologies, the TCAM array using *a*-IGZO FG transistors exhibits at least a 240× improvement in array-size scalability and a 2.7-fold reduction in search energy.

# Results

## Device structure and electrical characteristics of *a*-IGZO FG transistors

Figure 1a shows the schematic illustration of the fabricated *a*-IGZO FG transistor, and the corresponding scanning electron microscope (SEM) image is shown in Supplementary Fig. 1. The bottom-gate structure with top-contacted source/drain was used, where the gap between the source and drain is defined as the $L_{CH}$. Figure 1b shows the cross-sectional view of the device. High-quality *a*-IGZO with a thickness of ~8 nm and a small root-mean-square (RMS) roughness of ~0.55 nm was employed (roughness is shown in Supplementary Fig. 2). The gate stack consists of 6-nm $HfO_2$/20-nm TiN/10-nm $HfO_2$/30-nm TiN from top to bottom, where the inserted 20-nm TiN is an FG and can be used to trap charges. The thicknesses of each layer in the channel region (A-A in Fig. 1b) were confirmed by the high-resolution transmission electron microscopy (HRTEM) image (Fig. 1c). The smooth and sharp *a*-IGZO/$HfO_2$ interface and the amorphous nature of the IGZO layer were also demonstrated. Figure 1d-j illustrate the energy-dispersive X-ray (EDX) mapping profiles in the layer stack of the drain region (B-B in Fig. 1b). The distribution of Hf, Ti, N, In, Ga, Zn, and O elements are exhibited. It is noted that nitrogen (Fig. 1f) and oxygen (Fig. 1j) are present in all layers, suggesting the possible interlayer diffusion of N and O during the thermal annealing process.

Figure 2a shows the width-normalized $I_{DS}$-$V_{GS}$ curves of the fabricated *a*-IGZO FG transistor ($L_{CH}$ = 5 $\mu$m) at a low $V_{DS}$ of 0.1 V and a high $V_{DS}$ of 1 V. The transfer characteristics in the programmed state, which means the logic state '1', were measured by the direct-current (DC) method after applying a program pulse (6.5 V, 1 ms). Similarly, the transfer characteristics in the erased state, which indicates the logic '0', were measured after applying an erase pulse (-5 V, 1 ms). The electrons trapped at the FG shift the $V_{TH}$ of the transistor, which can be tuned by applied pulses because of the electron tunneling effect. To further illustrate the tunneling mechanism, the energy band diagram along the direction perpendicular to the channel was obtained by simulating an *a*-IGZO FG transistor with the same structure of our fabricated device using technology computer-aided design (TCAD). The simulated device structure is shown in Supplementary Fig. 3a. During programming, electrons in the *a*-IGZO channel tunnel into the FG through the tunneling oxide after applying a positive voltage on the control gate (Supplementary Fig. 3b). On the other hand, electrons tunnel back from the FG to the IGZO channel after performing a pulse with a negative voltage during erasing (Supplementary Fig. 3c). From the $I_{DS}$-$V_{GS}$ curves, the MW defined as the difference in $V_{TH}$ between the two states is ~1.8 V, achieving excellent memory characteristics in a low operating voltage. It should be noted that the $I_{OFF}$ of ~$10^{-7}$ $\mu$A/$\mu$m (~1 pA without normalization to the width of the device) is due to the detection limit of the measurement tool. The actual $I_{OFF}$ should be less than 1 pA for both programmed and erased states because of the wide bandgap of *a*-IGZO (~3.2 eV) and the low gate leakage current[52,55]. A large $I_{ON}/I_{OFF}$ of ~$1\times10^{8}$ is achieved at $V_{DS}$ of 1 V, and it can be further boosted by scaling the device $L_{CH}$ to improve $I_{ON}$.

*SS* values extracted from the $I_{DS}$-$V_{GS}$ curves are shown in Fig. 2b as a function of $I_{DS}$ for the programmed state and the erased state at low and high $V_{DS}$. All the *SS* values remain less than 100 mV/decade over more than 2 orders of $I_{DS}$ with the smallest value of ~80 mV/decade, indicating the excellent gate stack quality. In Fig. 2c, the plot of transconductance ($G_m$) at $V_{DS}$ = 1 V as a function of $V_{GS}$ exhibits a peak value of ~3.5 $\mu$S/$\mu$m for the programmed state and 3.2 $\mu$S/$\mu$m for the erased state. It also shows a noticeable window between the two states. The capacitance-voltage (*C*-*V*) curves were measured using the measurement configuration described in Fig. 2d and were utilized to extract the effective mobility ($\mu_{eff}$) of *a*-IGZO FG transistors using the equation (1):

$$\mu_{eff} = \frac{L_{CH} \times I_{DS}}{W_{CH} \times Q \times V_{DS}} . \quad (1)$$

Here $Q$ is the carrier charge density and $W_{CH}$ is the channel width. This is a widely used method in the CMOS device community. Figure 2e plots the *C*-*V* results at various frequencies ranging from 100 Hz to 1000 kHz. The carrier charge density was obtained by integrating the *C*-*V* curve. The extracted $\mu_{eff}$ as a function of carrier density ($N_{carrier}$) is shown in Fig. 2f. $\mu_{eff}$ of 32.6 $cm^2/V{\cdot}s$, which is higher than poly-Si in 3D NAND[13], and relatively high for oxide semiconductor materials, is achieved at $N_{carrier}$ of ~$3.8{\times}10^{12}$ $cm^{-2}$.

**Impact of channel length down-scaling**

The continuous improvement in the areal density of memory devices lies in the scaling of a single transistor, especially the channel length of the transistor. In order to investigate the impact of $L_{CH}$ down-scaling on device performance, transistors with various $L_{CH}$ were fabricated. The shortest $L_{CH}$ is 60 nm, which was confirmed by the SEM image (Supplementary Fig. 4). Figure 3a shows the transfer characteristics of the ultra-scaled transistor ($L_{CH}$ = 60 nm) with a $V_{DS}$ of 0.1 V and 1 V, where the $I_{ON}$ is significantly improved compared with that of the long-channel device ($L_{CH}$ = 5 $\mu$m) while keeping the $I_{OFF}$ below 1 pA. The MW is larger than 1.5 V. The extracted $G_m$ is plotted in Fig. 3b, showing a peak value of 22.3 $\mu$S/$\mu$m and 23.4 $\mu$S/$\mu$m for the programmed state and erased state, respectively. Figure 3c gives the output characteristics of the same device with an $L_{CH}$ of 60 nm, yielding a high current of ~127 $\mu$A/$\mu$m at $V_{DS}$ of 3 V and gate overdrive ($V_{GS}$ - $V_{TH}$) of 4 V.

The transfer characteristics of devices with $L_{CH}$ of 100 nm, 200 nm, and 500 nm are shown in Supplementary Fig. 5. Figure 3d summarizes the extracted *SS* as a function of $L_{CH}$ from the $I_{DS}$-$V_{GS}$ curves. The average *SS* of the programmed state and erased state remains less than 100 mV/decade until $L_{CH}$ is scaled down to 200 nm. The ultra-scaled device with an $L_{CH}$ of 60 nm exhibits the *SS* value of ~105 mV/decade, indicating excellent control of the SCEs. Summarized $I_{ON}$ can be observed in Fig. 3e, where $I_{ON}$ is obtained at an overdrive voltage of 3 V and a $V_{DS}$ of 3 V from the output characteristics (Supplementary Fig. 6). As the $L_{CH}$ scales down, the resistance in the channel region is reduced. However, the source/drain series resistance ($R_{SD}$) remains unchanged, and the $R_{SD}$ dominates the output current, which is the reason for the saturation of $I_{ON}$ at sub-500 nm of $L_{CH}$. By further reducing the $R_{SD}$ through advanced source/drain engineering, a significant $I_{ON}$ enhancement is expected[56,57]. Figure 3f shows the relationship between the MW and $L_{CH}$. It is noteworthy that the MW remains larger than 1.5 V with $L_{CH}$ down to 60 nm, demonstrating the promise of our *a*-IGZO FG transistors to be used in ultra-high density 3D NAND at advanced technology nodes.

**Pulse response and reliability of *a*-IGZO FG transistors**

The pulse response of a single *a*-IGZO FG transistor was studied by applying pulses with different amplitudes and widths. For the positive $V_{TH}$ shift by applying a program pulse, the device was initially set to an erased state using a pulse with an amplitude of -5 V and a width of 1 ms. After that, a positive pulse was applied to switch the device to the programmed state. The value of $V_{TH}$ was obtained by DC measurement after each pulse. For the case of erasing, a positive program pulse (6.5 V, 1 ms) and a negative erase pulse were employed, which was opposite to the programming case. Figure 4a and Figure 4b illustrate the 2D contour of the $V_{TH}$ shift as a function of pulse amplitude and pulse width for the program pulse and the erase pulse, respectively. The absolute value of the $V_{TH}$ shift increases with the increase of pulse amplitude and pulse width, and a pulse with 1 ms pulse width is sufficient to obtain a saturated $V_{TH}$ shift. Notably, the *a*-IGZO FG transistor requires a low operating voltage of 6 V, which is beneficial for power consumption reduction.

The endurance of the *a*-IGZO FG transistor was evaluated by 1000 cycles of continual switching between programmed and erased states with a $V_{DS}$ of 0.1 V, as depicted in Fig. 5a. Each transfer curve was measured after the program or erase pulse. The device has a width of 10 $\mu$m. Figure 5b summarizes the extracted voltage for all the curves from Fig. 5a at the current level of $10^{-5}$ $\mu$A/$\mu$m. It shows that both the programmed and erased $I_{DS}$-$V_{GS}$ curves are positively shifted, with the MW slightly reducing from 1.8 to 1.6 V after 1000 cycles. The shift of the curves may be caused by the asymmetry of the program pulse and erase pulse, where the electrons were not completely depleted from the FG by the erase pulse and accumulated after cycling. Figure 5c shows the retention characteristics of the device after applying a program pulse (6.5 V, 1 ms) and an erase pulse (-5 V, 1 ms). Extrapolation of the data indicates that the programmed state and erased state can still be identified after 10 years with an MW of 0.9 V, demonstrating a retention time longer than 10 years. The change of $V_{TH}$ in the programmed state is slightly faster than that of the erased state. It could be caused by the local trap-assisted tunneling, which stimulates electrons to tunnel from FG back to the channel[58].

Temperature-dependent operations of *a*-IGZO FG transistors were also investigated. The transfer characteristics at various temperatures from -40 to 80 °C with a step of 20 °C are described in Supplementary Fig. 7. *SS* increases monotonically as the temperature increases and reaches around 100 mV/decade at 80 °C, as depicted in Fig. 5d. The current obtained at $V_{GS}$ of 3

V and $V_{DS}$ of 0.1 V at various temperatures is shown in Fig. 5e. Improved current at higher temperatures could be explained by the increase of carrier concentration. This is an advantage for operations at higher temperatures over Si-based devices, where a reduction of current at higher temperatures was typically observed because of the mobility degradation caused by phonon scattering[59,60]. Figure 5f plots the extracted voltage and MW as a function of the operating temperature. The $I_{DS}$-$V_{GS}$ curves exhibit a slight positive shift when the operating temperature increases from -40 °C to 20 °C. However, further increasing the operating temperature from 20 to 80 °C causes the curves to shift negatively, which could be dominated by the increased carrier concentration in the IGZO channel at high temperatures. It was observed that MW degrades slightly at high temperatures but still offers a value of ~1.4 V at 80 °C.

The performance benchmarking of our devices with reported flash memories based on an *a*-IGZO channel was summarized in Supplementary Table 1. The devices studied in this work have the shortest $L_{CH}$ of 60 nm with the best control of SCEs, achieving the highest $I_{ON}$ of 127 $\mu$A/$\mu$m and the lowest operating voltage among all the *a*-IGZO-based flash devices.

**Fully functional TCAM based on *a*-IGZO FG transistors**

The conventional CMOS-based TCAM cell is composed of 2 static random-access memory (SRAM) cells and 4 comparison transistors (Fig. 6a). The logic state of the TCAM cell is indicated by the states of SRAM cells, logically taking one of the three states: 0, 1, or X (don't care). The search operations are performed by 2 search lines (SL and $\overline{SL}$) while the write operations are performed by the bit lines and the write line. The output of the match line (ML) is determined by the XNOR result of the TCAM state and the search bit (Supplementary Table 2). It is clear that CMOS-based TCAM has complicated connection lines and occupies a large area.

Alternatively, the *a*-IGZO FG transistor-based TCAM is much more area-efficient due to its reduced connection lines, where the bit line and search line are merged into a single search line, as shown in Fig. 6b. This area-efficient schematic utilizes the nonvolatility and intrinsic transistor nature of the *a*-IGZO FG device, where the $V_{TH}$ can be tuned by pulses to indicate different logic states.

The SEM image in Supplementary Fig. 8 shows the fabricated TCAM cell with two parallel-connected *a*-IGZO FG transistors ($L_{CH}$ = 5 $\mu$m). The source terminals and drain terminals of the transistors are connected and defined as ML and ground, respectively, while the gate terminals are designed as the search lines. The transistor $T_0$ and $T_1$ in the TCAM cell can be set to logic states '1' or '0' after writing pulses (1, high $V_{TH}$ and 0, low $V_{TH}$), as summarized in Fig. 6c. The 'don't care' state can be realized by setting both transistors to high $V_{TH}$ states. For search operations, the pulse width is 0.2 ms with a low search voltage ($V_{SL_L}$) of 0 V or a high search voltage ($V_{SL_H}$) of 0.9 V. It should be noted that the negative erase pulse (-4.2 V, 1 ms) only partially shifts the $V_{TH}$ of the transistor in order to obtain a high current ratio between the search voltage of 0.9 V and 0 V ($I_{DS}$-$V_{GS}$ curves after pulses shown in Supplementary Fig. 9). This is because when the transistor is entirely erased, it can be switched on even at the search voltage of 0 V, or another method is to apply a lower $V_{SL_L}$, such as -1 V[41]. Note that the search operation is simply a transistor read, therefore its speed reported here is by no means the limit of the *a*-IGZO flash transistors, but rather limited by the measurement setup. Figure 6b describes the measurement setup of the TCAM cell with a $V_{DD}$ of 1.2 V and a load resistor of 3 MΩ. More details of the measurement scheme are described in Supplementary Fig. 10. During cell operation, logic states are first written into $T_0$ and $T_1$, and if the search state matches the stored data, the ML keeps high; otherwise, the ML discharges and switches to low.

Figure 6d shows the operation scheme of writing ‘1’ of the TCAM cell, where $T_0$ is set to ‘1’ with a high $V_{TH}$ after applying a positive pulse, and $T_1$ is set to ‘0’ with a low $V_{TH}$. After that, search schemes of searching ‘1’ and searching ‘0’ are performed, as depicted in Fig. 6e and Fig. 6f, respectively. In Fig. 6e, a search pulse with an amplitude of $V_{SL_H}$ is applied to $T_0$, and another search pulse with an amplitude of $V_{SL_L}$ is applied to $T_1$. $T_0$ with a high $V_{TH}$ is cut off when the search voltage is 0.9 V, and $T_1$ with a low $V_{TH}$ is also cut off when the search voltage is 0 V. Therefore, the ML remains high, meaning that the search data matches with the data in the cell. On the other hand, a mismatch occurs when operating a search ‘0’ scheme. $T_0$ with a high $V_{TH}$ is still off under the search voltage of 0 V, but $T_1$ with a low $V_{TH}$ is switched on under the search voltage of 0.9 V, which discharges the ML, as observed in Fig. 6f. Other cases of writing ‘0’ and ‘X’ and corresponding search schemes are shown in Supplementary Fig. 11. Thus, a fully functional TCAM cell is demonstrated.

The ultra-low $I_{OFF}$ of *a*-IGZO FG transistors (less than $10^{-7}$ $\mu$A/$\mu$m for *a*-IGZO FG transistors, while ~$10^{-4}$ $\mu$A/$\mu$m for Si transistors[61,62]) also provides the opportunity to make a TCAM array with a large array size, which can be used for hyperdimensional computing. To compare the performance of *a*-IGZO FG transistor-based TCAM with various TCAM designs (16T-CMOS, 2T2R, and 2FeFET), array-level simulations were carried out using the framework shown in Fig. 7a[63,64]. TCAM cell for simulation is well-calibrated with the experimental data of our fabricated *a*-IGZO FG transistor with an $L_{CH}$ of 60 nm, as shown in Supplementary Fig. 12. $V_{SL_L}$ is set to -2 V while the $V_{SL_H}$ is set to vary from 0 to 1 V to study the impact of $I_{OFF}$ on the array size. During simulations, the ML of each row is precharged to a high voltage level, and the search lines are driven with input data. The sense amplifier senses the ML state of the associated word. When at least one cell does not match the input data, the ML could be discharged and drop to a low level. Only when all cells match with the input data, the ML remains high. However, as the array size expands, the total $I_{OFF}$ increases because more TCAM cells are located at a single row. The ML could be discharged even at the matching case, making the matching case unable to be maintained for a long time. Figure 7b plots the simulation results of the matching case (storing 0 and searching 0) for *a*-IGZO FG transistor-based TCAM rows with various sizes ($V_{SL_L}$ = -2 V, $V_{SL_H}$ = 0 V). The actual search delay should be at the level of ns. However, to illustrate the ML dropping clearly, the search time is extended to 100 $\mu$s here. Obviously, the voltage of the ML for the 1×64 array drops faster than that of the 1×1 array because of the increased total $I_{OFF}$.

Time to drop $V_{DD}$/2 at the matching case, which is utilized to evaluate the scalability, is summarized in Fig. 7c for various TCAM designs based on a 1×64 TCAM array. TCAM array using *a*-IGZO FG transistors ($V_{SL_L}$ = -2 V, $V_{SL_H}$ = 0 V) shows at least a 240× improvement in time to drop to $V_{DD}$/2 as compared with 16T-CMOS, 2T2R, and 2FeFET TCAMs, indicating excellent scalability and the potential to be used in an enormous TCAM array which is essential for many emerging applications. It is important to mention that the $I_{OFF}$ of *a*-IGZO FG transistors of ~1 pA was obtained from the experimental measurement and limited by the detection accuracy of the measurement tool. The actual $I_{OFF}$ should be much lower so that the time to drop to $V_{DD}$/2 of the *a*-IGZO FG transistor-based TCAM array can be much longer. Additionally, the search energy and search delay of a 64×64 TCAM array are also simulated, as depicted in Fig. 7d. The *a*-IGZO FG transistor-based TCAM ($V_{SL_L}$ = -2 V, $V_{SL_H}$ = 0 V) array shows at least a 2.7-fold reduction in search energy due to the low $I_{OFF}$ and simple cell structure. The search delay defined as the time to drop to $V_{DD}$/2 of the ML under the mismatching case is mainly determined by the turn-on current of the transistor with a low $V_{TH}$ state under $V_{SL_H}$ (Supplementary Fig. 12). Higher current discharges the ML faster and results in a lower search delay. By increasing the $V_{SL_H}$, the discharging current can be enhanced, and the search delay can be reduced while keeping the search energy almost unchanged (Fig. 7d). However, there is a trade-off between the search delay and the scalability (Fig. 7c). The time to drop to $V_{DD}/2$ at the matching case is reduced because the $I_{OFF}$ increases with the increase of $V_{SL_H}$, leading to degraded scalability. This can be optimized in future work by lowering the $R_{SD}$ of transistors to boost the discharging current and to remain the $V_{SL_H}$ at 0 V.

## Conclusion

In summary, 3D NAND compatible high-performance *a*-IGZO FG transistors and an area-efficient TCAM cell based on two *a*-IGZO FG transistors are fabricated and investigated in this paper with both experiments and simulations. Our *a*-IGZO FG transistors have the benefits of low process temperature, low operating voltage, less than $10^{-7}$ $\mu$A/$\mu$m $I_{OFF}$, large $I_{ON}/I_{OFF}$ of more than $1\times10^{8}$, low *SS* of about 80 mV/decade, high effective mobility of 32.6 $cm^2/V\cdot s$, large MW of ~1.8 V (1.4 V at 80 °C), and good endurance and retention. Ultra-scaled *a*-IGZO FG transistor with an $L_{CH}$ of 60 nm offers a boosted $I_{ON}$ of 127 $\mu$A/$\mu$m while remaining the MW larger than 1.5 V. Attributing the low $I_{OFF}$ of *a*-IGZO FG transistor and the simple TCAM structure, the TCAM array based on *a*-IGZO FG transistors shows much better scalability and lower search energy as compared with TCAM arrays based on 16T-CMOS, 2T2R, and 2FeFET. This work paves the way for the applications of *a*-IGZO FG transistors in 3D NAND and TCAM and projects the potential to significantly expand the size of the TCAM array.

## Methods

**Device fabrication.**

The device fabrication process (Supplementary Fig. 13) started with the deposition of 30 nm TiN on the $SiO_2$/Si substrate using e-beam evaporation followed by an etching process. 10 nm $HfO_2$ film was deposited by ALD at 250 °C. $Hf[N(C_2H_5)CH_3]_4$ (TEMAHf) and ozone were used as the Hf precursor and oxygen source, respectively. After the patterning and etching of $HfO_2$, another 20 nm TiN was deposited and patterned as the floating gate. Then, 6 nm $HfO_2$ deposited by ALD was performed as the tunneling layer. After etching the tunneling layer, a layer of 8-nm *a*-IGZO was deposited by radio frequency (RF) sputtering at room temperature. Device isolation was then performed by wet etching of IGZO using HCl solution. Source/drain regions were defined using electron beam lithography (EBL), and Ti/Pt electrodes were deposited by e-beam evaporation followed by the lift-off process. Finally, fabricated devices were annealed at 300 °C for 20 minutes. The TCAM cell was formed during source/drain patterning by connecting the source and drain terminals of two transistors, respectively.

**Device characterization.**

Temperature-dependent operations of *a*-IGZO FG transistors were measured under vacuum conditions at various temperatures, and other characteristics of devices were measured under air conditions and room temperature. The surface roughness was measured by atomic force microscopy (AFM) (NX20, Park). The electrical characteristics of *a*-IGZO FG transistors were measured by a semiconductor parameter analyzer (4200a-SCS, KEITHLEY), and pulse measurements were performed by the pulse measurement unit (4225-PMU, KEITHLEY). The write and search functions of the TCAM cell were carried out by the analyzer, oscilloscope (MDO3104, Tektronix), and DC power supply (E3630A, Agilent). The schematic of the measurement setup is provided in Supplementary Fig. 10. The ML of the TCAM cell was connected to an external resistor (3 MΩ). The other terminal of the resistor was connected to the power supply with a $V_{DD}$ of 1.2 V. Input pulses for writing and searching were generated by the PMU. The amplitude of the program pulse and erase pulse were set to 6.0 V and -4.2 V, respectively, with the pulse width of 1 ms. 200 $\mu$s search pulses with amplitudes of 0.9 V and 0 V were used for searching operations. The waveforms of the search lines and the ML were plotted by the oscilloscope.

## Author contributions

This project was supervised and directed by X. Gong. C. Sun and X. Gong conceived and designed the experiments. C. Sun performed device fabrication and electrical measurements. C. Li, K. Ni, and X. Yin carried out the array-level simulation. All authors contributed to the discussion and data analysis. C. Sun, K. Ni, and X. Gong wrote the manuscript.

## Competing interests

The authors declare no competing interests.

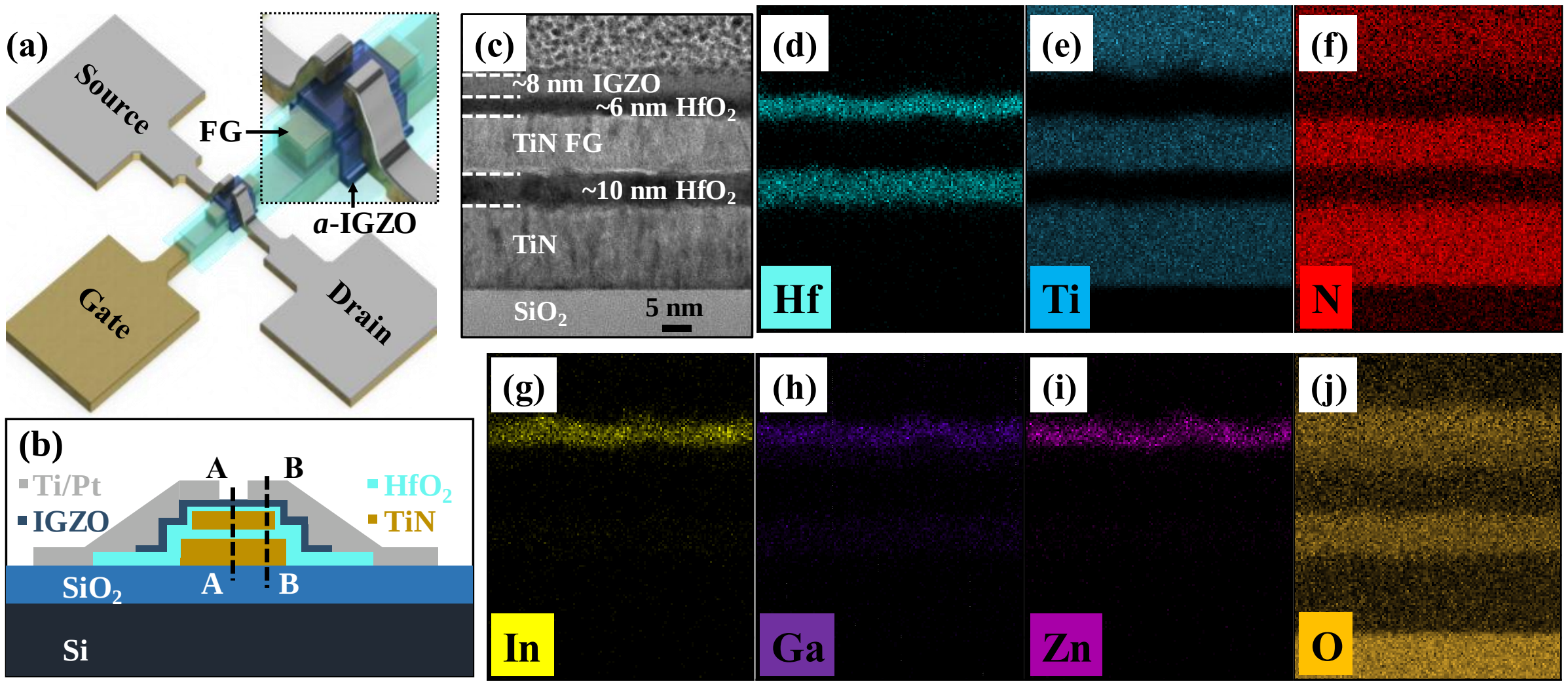

**Fig. 1 Device structure of the *a*-IGZO FG transistor. a** Schematic illustration of the *a*-IGZO FG transistor. $L_{CH}$ is defined as the gap between the source and drain. **b** Cross-sectional view of the device, employing a gate stack of $HfO_2$/TiN (FG)/$HfO_2$/TiN from top to bottom. **c** HRTEM image of the device in the channel region, confirming the thickness of each layer. **d-j**. EDX mapping profiles in the drain region, showing the distribution of Hf, Ti, N, In, Ga, Zn, and O elements.

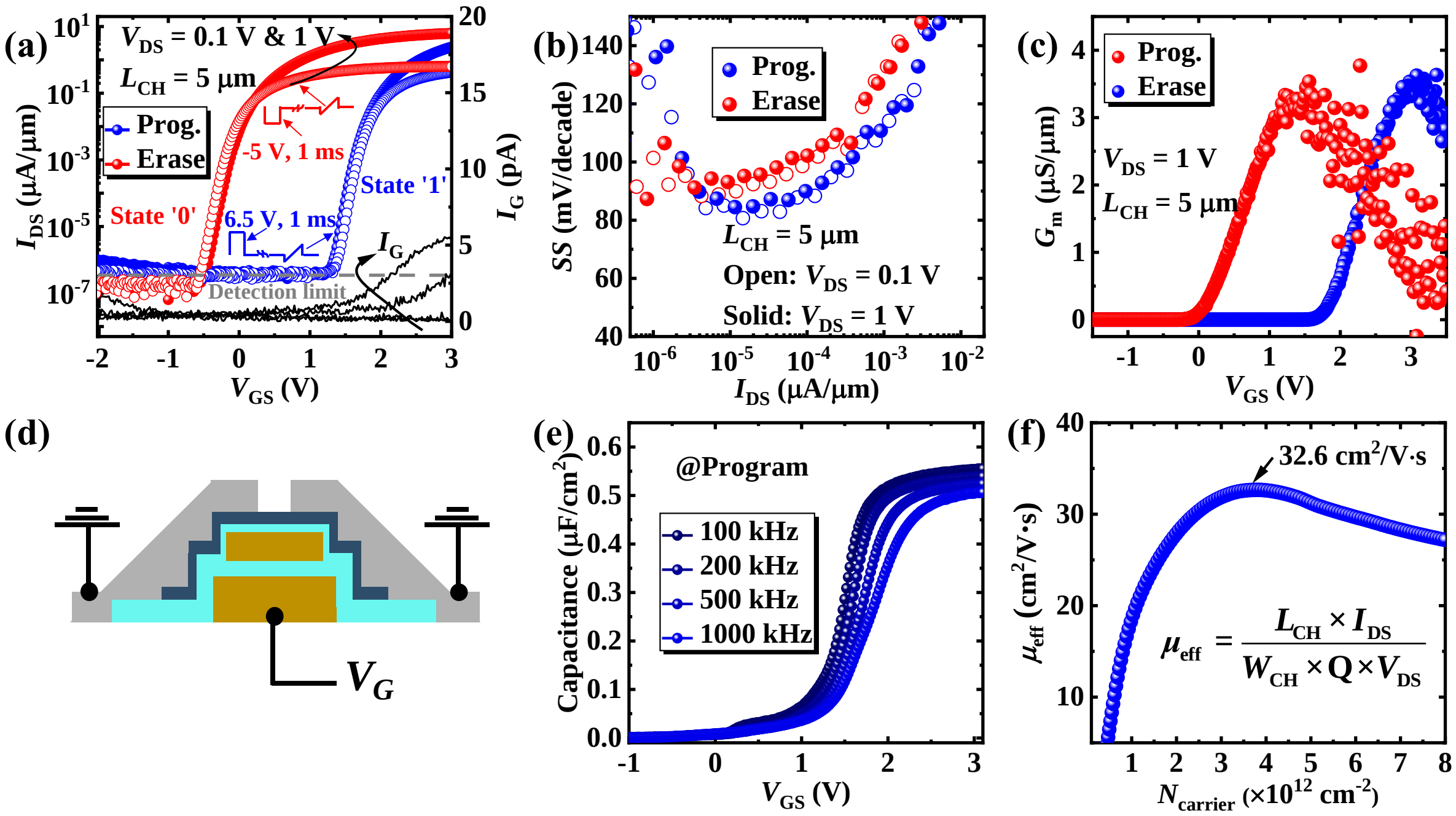

**Fig. 2 Electrical characterization of the *a*-IGZO FG transistor with an $L_{CH}$ of 5 $\mu$m. a** $I_{DS}$-$V_{GS}$ of the device measured after applying pulses, achieving a large MW of 1.8 V. $I_{OFF}$ is lower than the detection limit of the measurement tool. Programmed state means '1', and erased state means '0'. **b** Extracted *SS* from $I_{DS}$-$V_{GS}$ curves. All the *SS* values remain less than 100 mV/decade over more than 2 orders, with the smallest value of ~ 80 mV/decade. **c** Extracted $G_m$ from $I_{DS}$-$V_{GS}$ curves at $V_{DS}$ of 1 V, showing a noticeable window between the two states. **d** Configuration for *C*-*V* measurements. The source and the drain of the devices are connected. **e** *C*-*V* measurements of the device for mobility extraction. **f** Extracted mobility of the device using the inserted equation, achieving high mobility of 32.6 $cm^2/V\cdot s$.

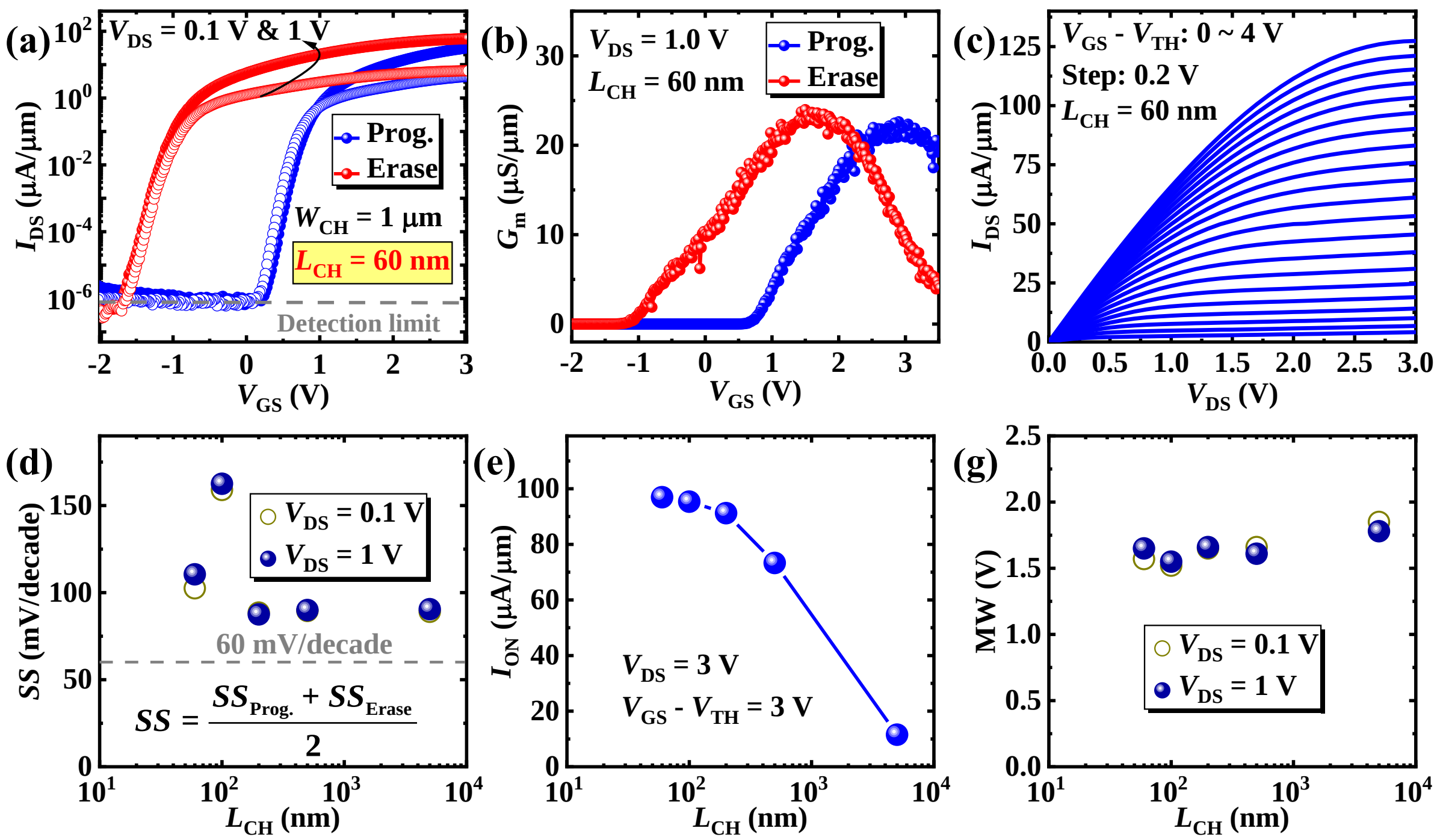

**Fig. 3 Impact of $L_{CH}$ scaling and characterization for *a*-IGZO FG transistors. a** $I_{DS}$-$V_{GS}$ of the ultra-scaled device with an $L_{CH}$ of 60 nm, showing a boosted $I_{ON}$ while keeping the $I_{OFF}$ below 1 pA. The MW is larger than 1.5 V for both $V_{DS}$ of 0.1 V and 1 V. **b** $G_m$ extracted from the $I_{DS}$-$V_{GS}$ curves with a peak value of 22.3 $\mu$S/$\mu$m for the programmed state and 23.4 $\mu$S/$\mu$m for the erased state. **c** $I_{DS}$-$V_{DS}$ curves of the device ($L_{CH}$ = 60 nm). A high current of ~127 $\mu$A/$\mu$m is obtained at $V_{DS}$ = 3 V and $V_{GS}$ - $V_{TH}$ = 4 V. **d** Impact of $L_{CH}$ on *SS*. *SS* slightly increases as the scaling of $L_{CH}$ with a value of ~105 mV/decade for $L_{CH}$ of 60 nm, indicating a good control of SCEs. **e** Summary of $I_{ON}$ at an overdrive voltage of 3 V and $V_{DS}$ of 3 V. **f** MW as a function of $L_{CH}$. MW maintains stable as the $L_{CH}$ reduces.

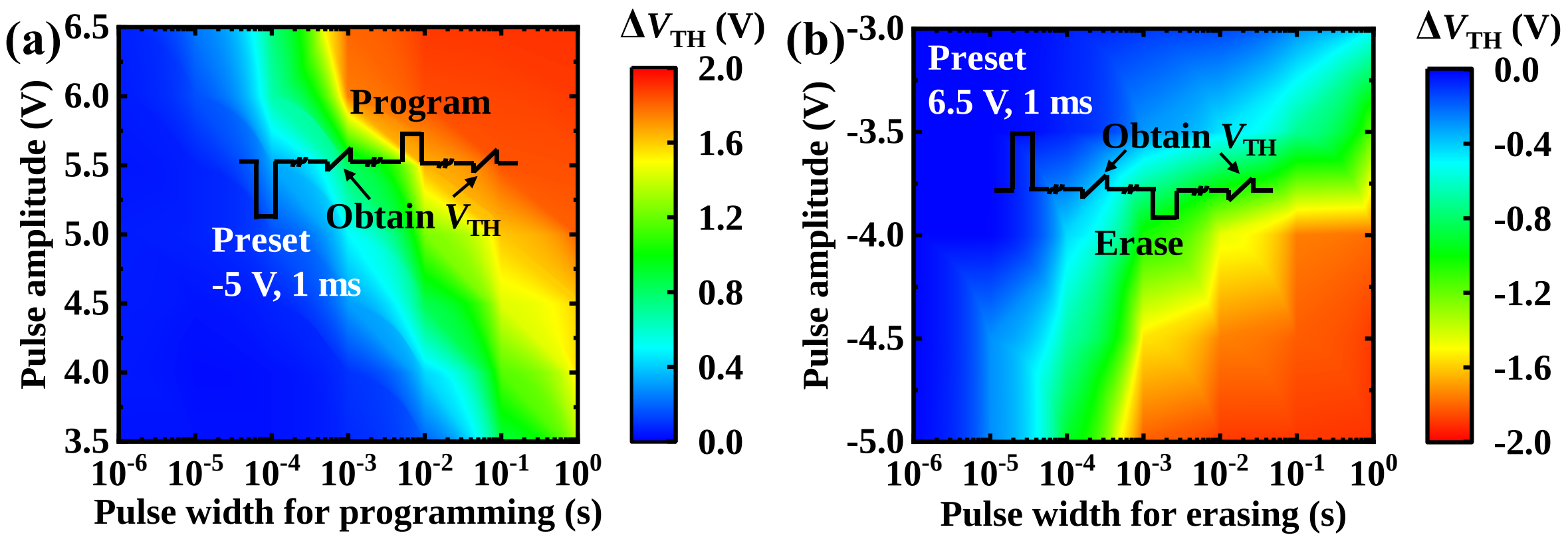

**Fig. 4 $V_{TH}$ tuning of *a*-IGZO FG transistors by pulses. a** $V_{TH}$ shifts positively by applying program pulses. The device was initially set to an erased state. The $V_{TH}$ was obtained by DC measurement after each pulse. Pulse with an amplitude of 6 V (1 ms) is sufficient to program the device entirely. **b** $V_{TH}$ shifts negatively by applying erase pulses. The device was initially set to a programmed state.

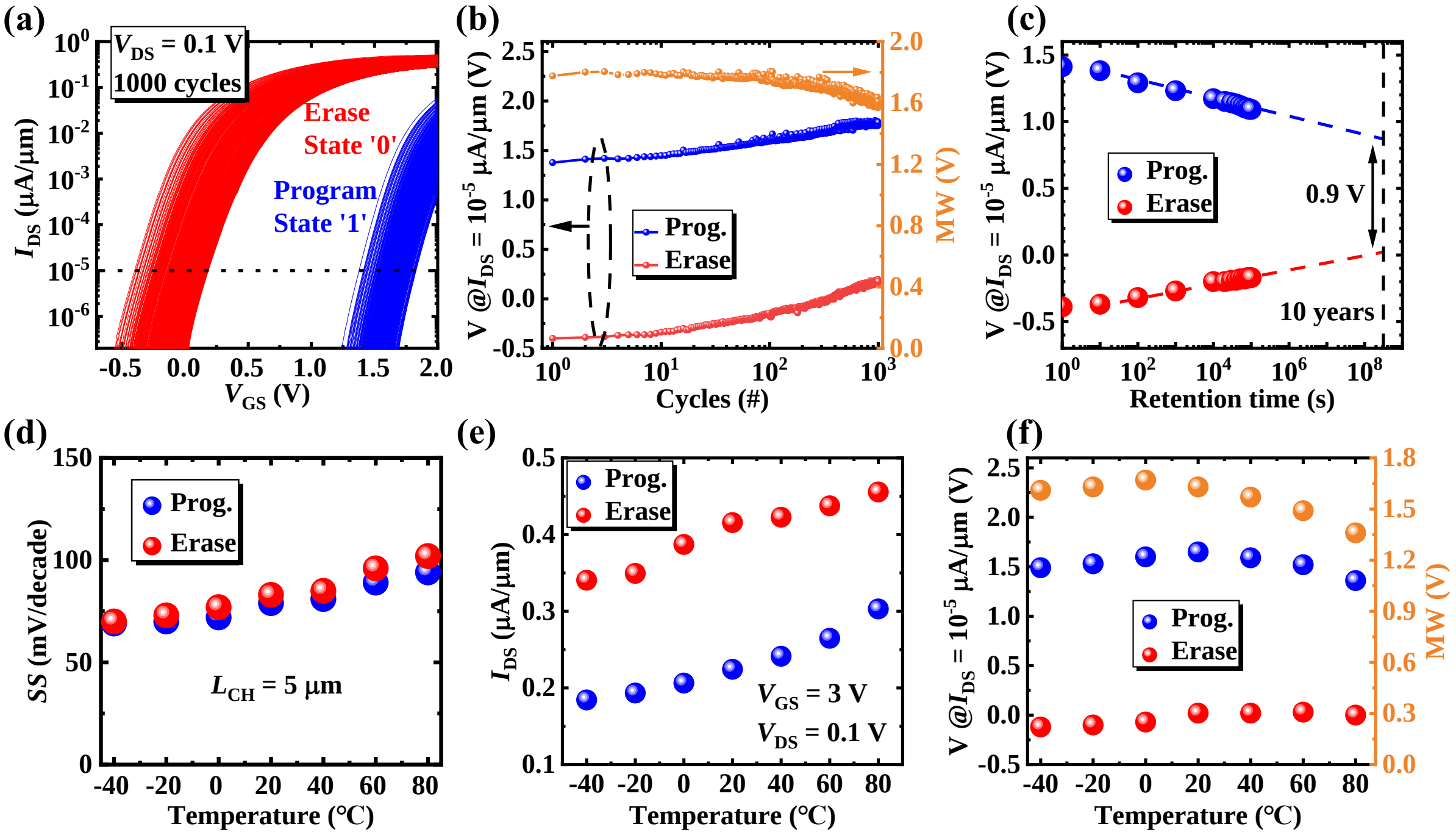

**Fig. 5 Reliability characteristics of *a*-IGZO FG transistors. a** Endurance measurement performed by 1000 cycles of continual switching between the programmed state and erased state. Each curve was obtained by DC measurement after applying the pulses. **b** Extracted voltage level at $I_{DS} = 10^{-5}$ $\mu$A/$\mu$m and the corresponding MW from endurance measurement. The positive shift of the voltage level could be caused by the asymmetry of the program pulse and erase pulse, resulting in the accumulation of electrons at the FG after 1000 cycles. The MW slightly drops to 1.6 V from 1.8 V after 1000 cycles. **c** Retention measurement of the device. $V_{TH}$ shows a positive shift after erasing and a negative shift after programming. Extrapolation of the points indicates that the device has a retention time longer than 10 years. **d** The relationship between *SS* and the operating temperature of the device. *SS* increases monotonically as the temperature increases. **e** Current at $V_{GS}$ = 3 V and $V_{DS}$ = 0.1 V increases monotonically as the temperature increases for programmed and erased states. **f** Extracted voltage level at $I_{DS} = 10^{-5}$ $\mu$A/$\mu$m from $I_{DS}$-$V_{GS}$ for various temperatures. The device is able to maintain an MW of ~1.4 V at 80 °C.

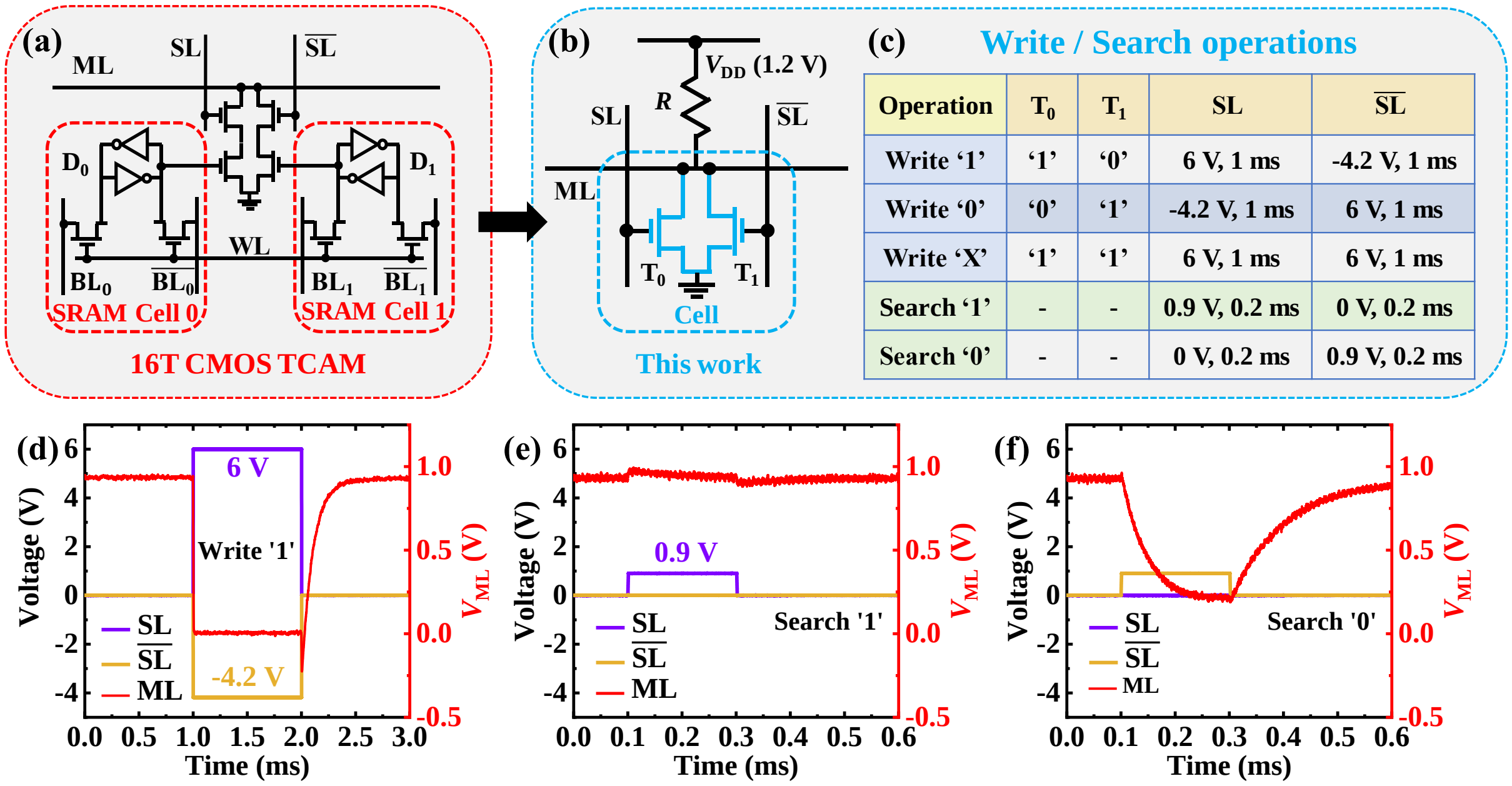

**Fig. 6 Fully functional TCAM cell using *a*-IGZO FG transistors. a** Schematic of a conventional CMOS-based TCAM cell, which is composed of 2 SRAM cells and 4 comparison transistors. **b** Schematic of the area-efficient TCAM cell based on *a*-IGZO FG transistors. SL and $\overline{SL}$ are employed for writing and searching. The load resistor (3 MΩ) is externally connected for measuring the voltage of ML. **c** Summary of write and search operations. The state of the TCAM cell is indicated by the states of $T_0$ and $T_1$. Pulses with amplitudes of 0.9 V and 0 V and a width of 0.2 ms are applied for searching. **d** Operation scheme of writing '1' into the TCAM cell by setting $T_0$ to '1' and $T_1$ to '0'. **e** Operation scheme of searching '1' after writing '1'. $T_0$ with high $V_{TH}$ is cut off when the search pulse is 0.9 V, while $T_1$ with low $V_{TH}$ is also cut off when the search pulse is 0 V, indicating a matching case and keeping the ML at a high voltage level. **f** Operation scheme of searching '0' after writing '1'. $T_0$ with high $V_{TH}$ is cut off when the search pulse is 0 V, but $T_1$ with low $V_{TH}$ is switched on when the search pulse is 0.9 V, indicating a mismatching case and discharging the ML.

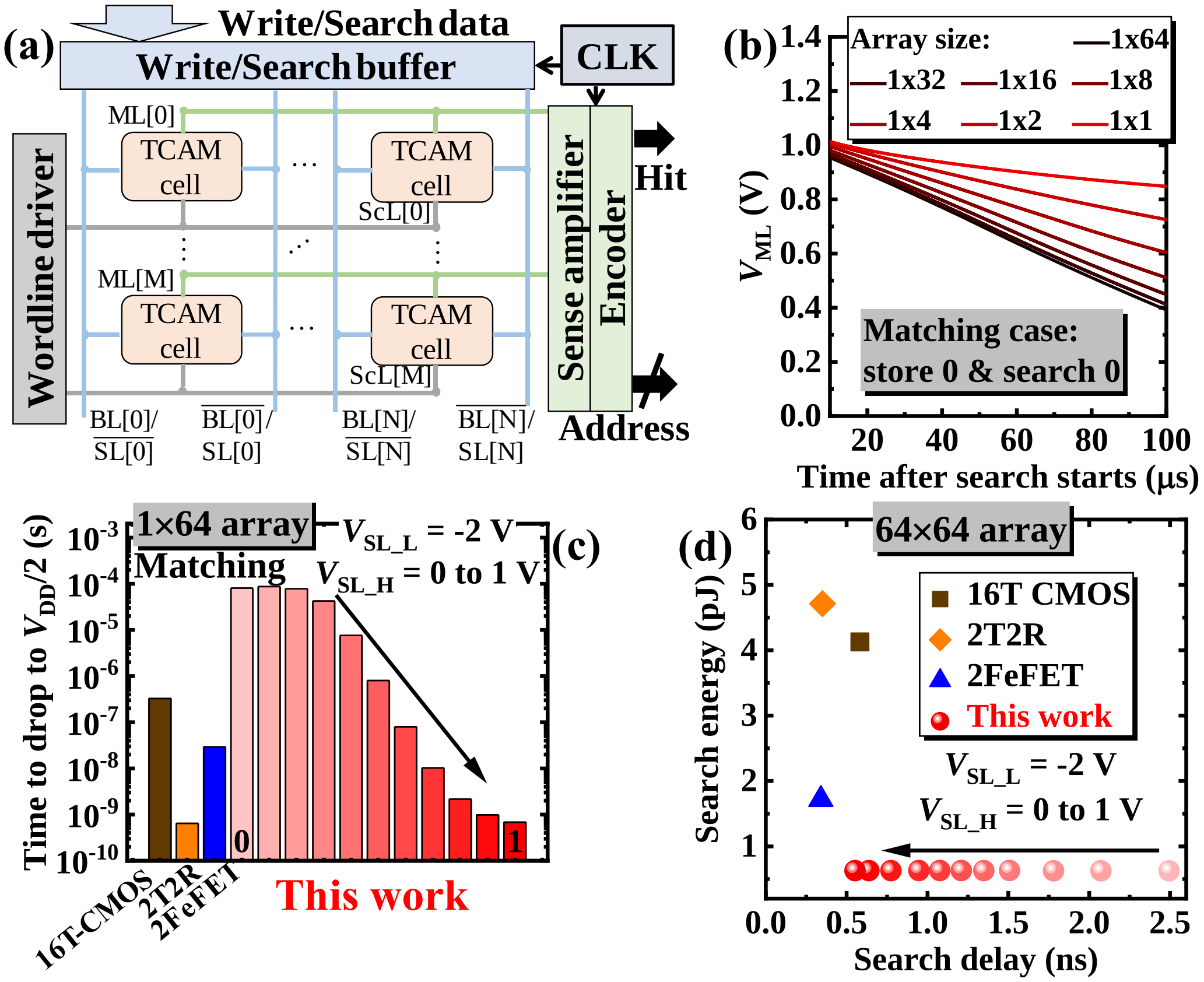

**Fig. 7 Array-level simulations for TCAMs. a** Simulation framework. The ML is precharged before data input. ML remains high only when all the cells match with the input data. **b** Waveforms of ML at the matching case. The ML is discharged by the total $I_{OFF}$ even all the transistors are cut off. Here the search time is extended to 100 $\mu$s to illustrate the ML dropping clearly. **c** Time to drop to $V_{DD}/2$ of the ML for TCAM arrays using various designs. TCAM array based on our *a*-IGZO FG transistors shows at least a 240× improvement in the time compared with 16T-CMOS, 2T2R, and 2FeFET TCAMs ($V_{SL_L}$ = -2 V, $V_{SL_H}$ = 0 V). **d** Benchmarking of the search energy and search delay for TCAM arrays. TCAM array based on *a*-IGZO FG transistors achieves at least a 2.7-fold reduction in search energy. The search delay can be reduced by increasing the $V_{SL_H}$ while keeping low search energy.

# Supplementary Information

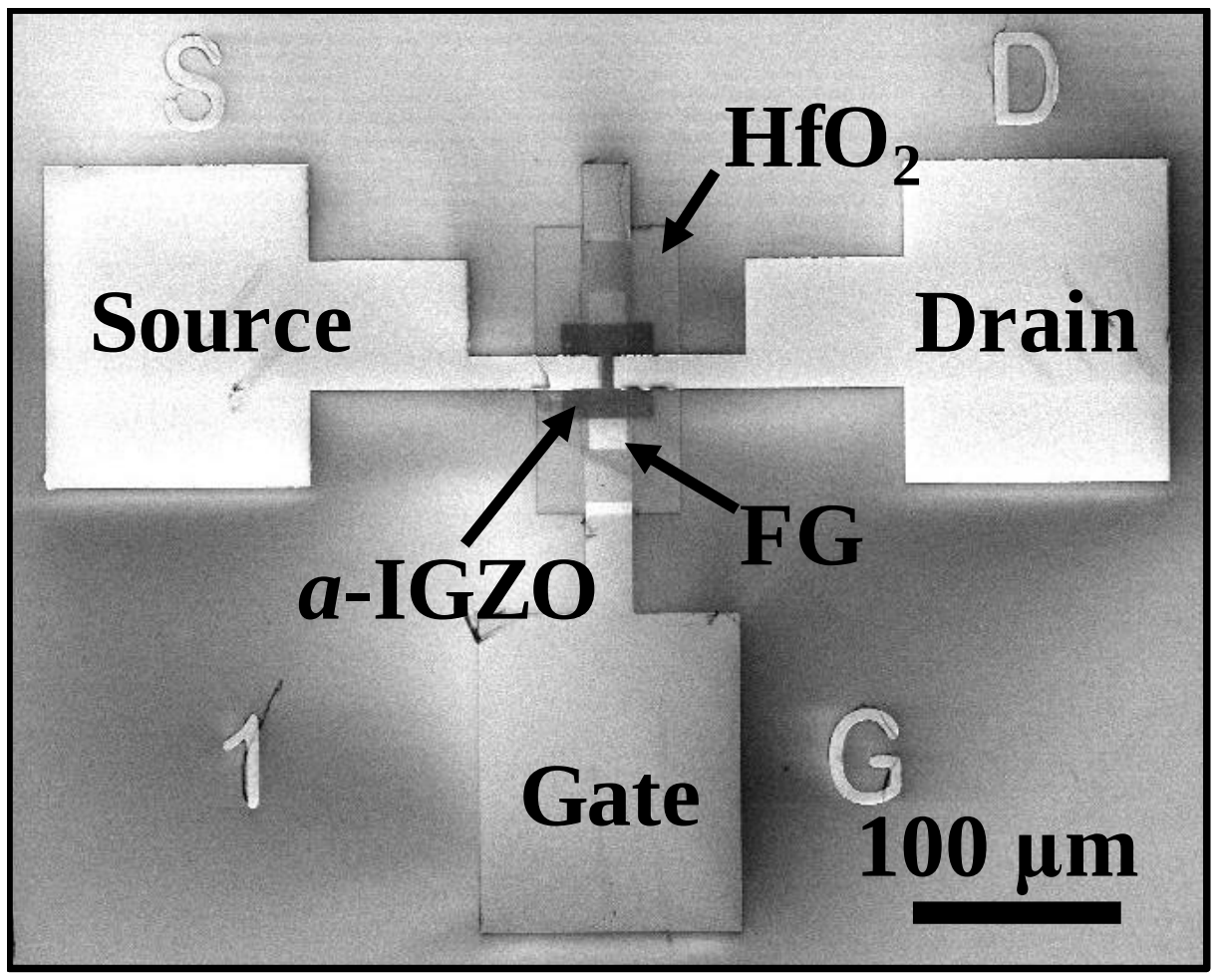

**Supplementary Fig. 1 Top-view SEM image of the FG *a*-IGZO TFT.** The gap between the source and drain is defined as the channel length ($L_{CH}$). Channel width ($W_{CH}$) is 10 $\mu$m, and $L_{CH}$ is 5 $\mu$m here.

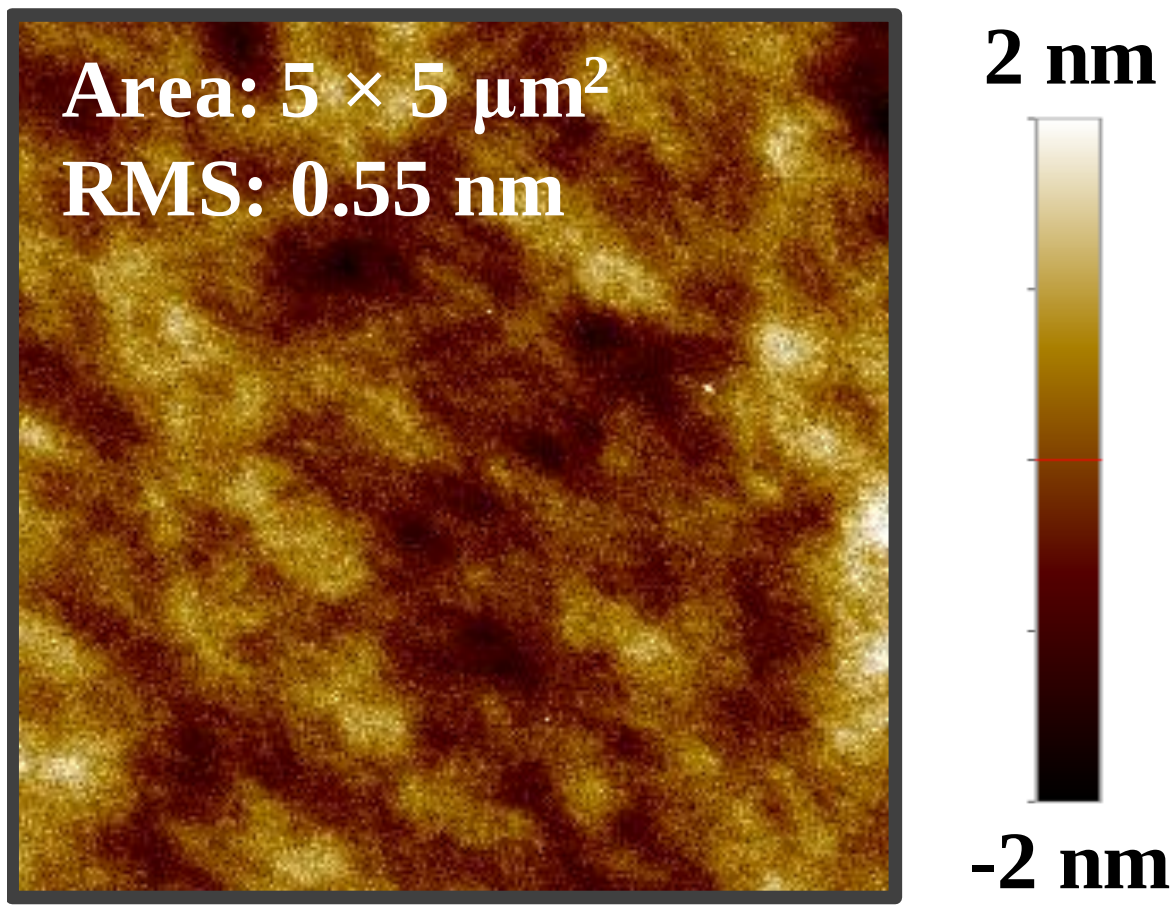

**Supplementary Fig. 2 AFM image of the *a*-IGZO layer.** *a*-IGZO with a thickness of ~8 nm shows a smooth surface with root-mean-square (RMS) roughness of 0.55 nm, which is confirmed by the AFM image with a scan area of 5 $\mu$m × 5 $\mu$m.

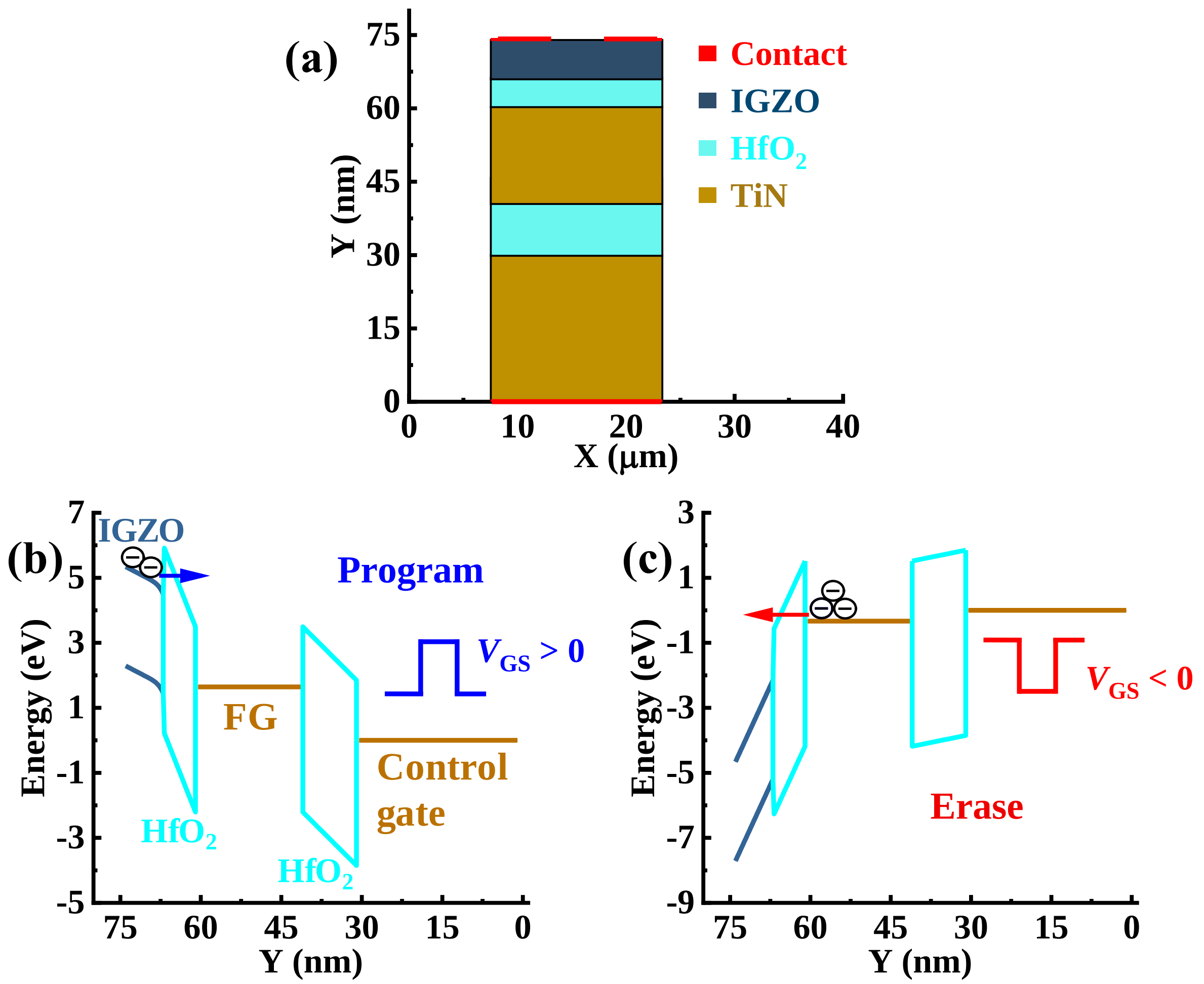

**Supplementary Fig. 3 Simulation of FG *a*-IGZO TFTs to illustrate the tunneling mechanism. a** Structure of the device. The source/drain metal is not drawn and replaced by the contact. Energy band diagrams are extracted along the direction perpendicular to the channel. **b** Energy band diagram of the device under a positive $V_{GS}$. Electrons in the *a*-IGZO channel tunnel into the FG through the tunneling oxide. **c** Energy band diagram of the device under a negative $V_{GS}$. Electrons tunnel back after applying a negative voltage on the control gate.

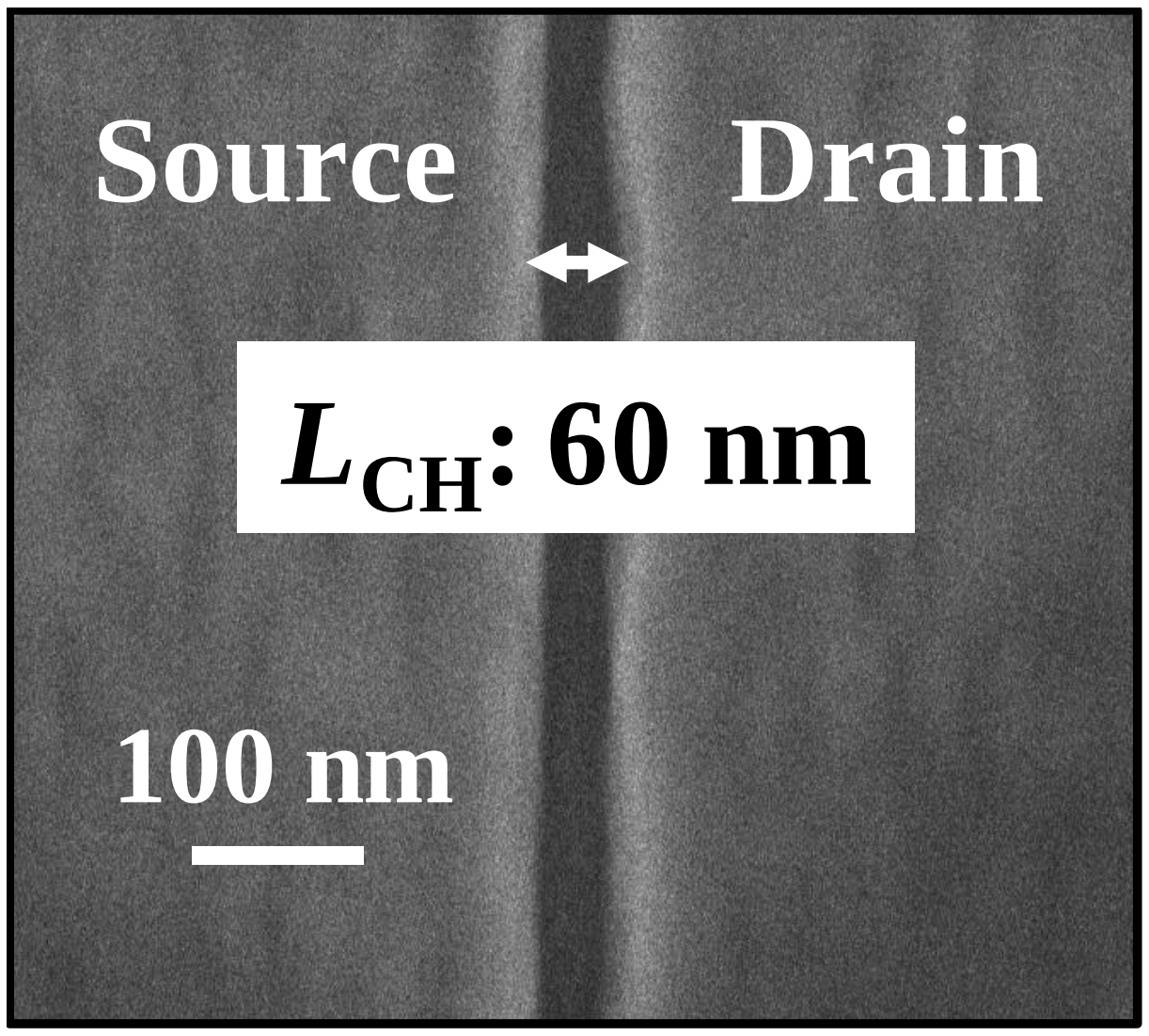

**Supplementary Fig. 4 Top-view SEM image of the ultra-scaled device channel.** The $L_{CH}$ of 60 nm is confirmed by the SEM image.

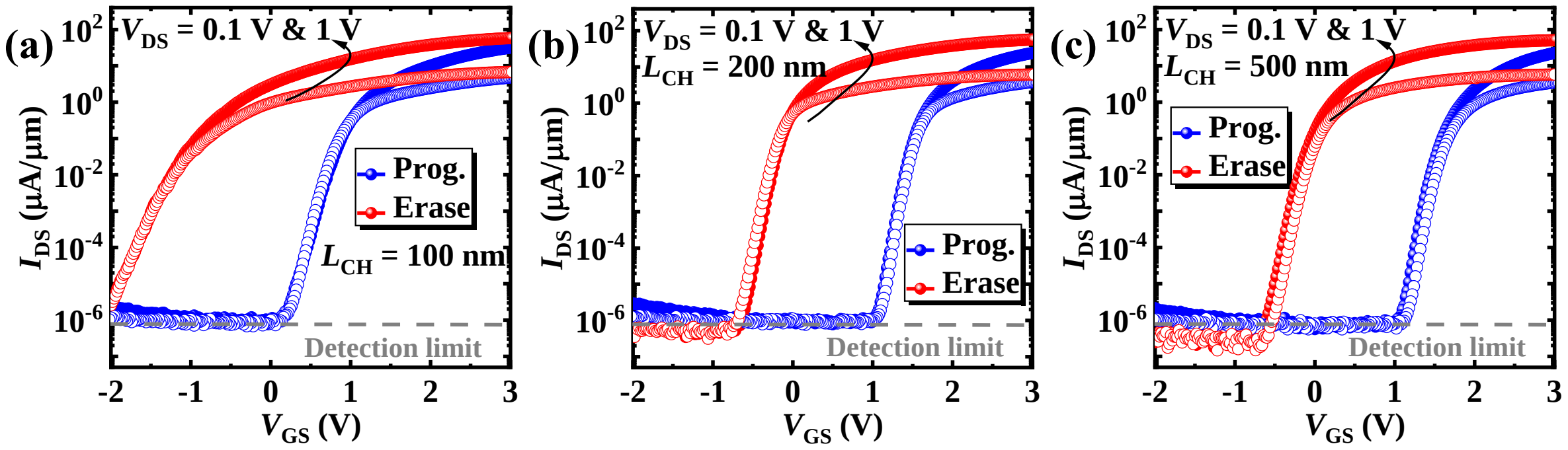

**Supplementary Fig. 5 Transfer characteristics of devices with different $L_{CH}$s.** Transfer characteristics of the device with $L_{CH}$ of (**a**) 100 nm, (**b**) 200 nm, and (**c**) 500 nm at $V_{DS}$ of 0.1 V and 1 V are obtained by DC measurement after applying program pulse or erase pulse.

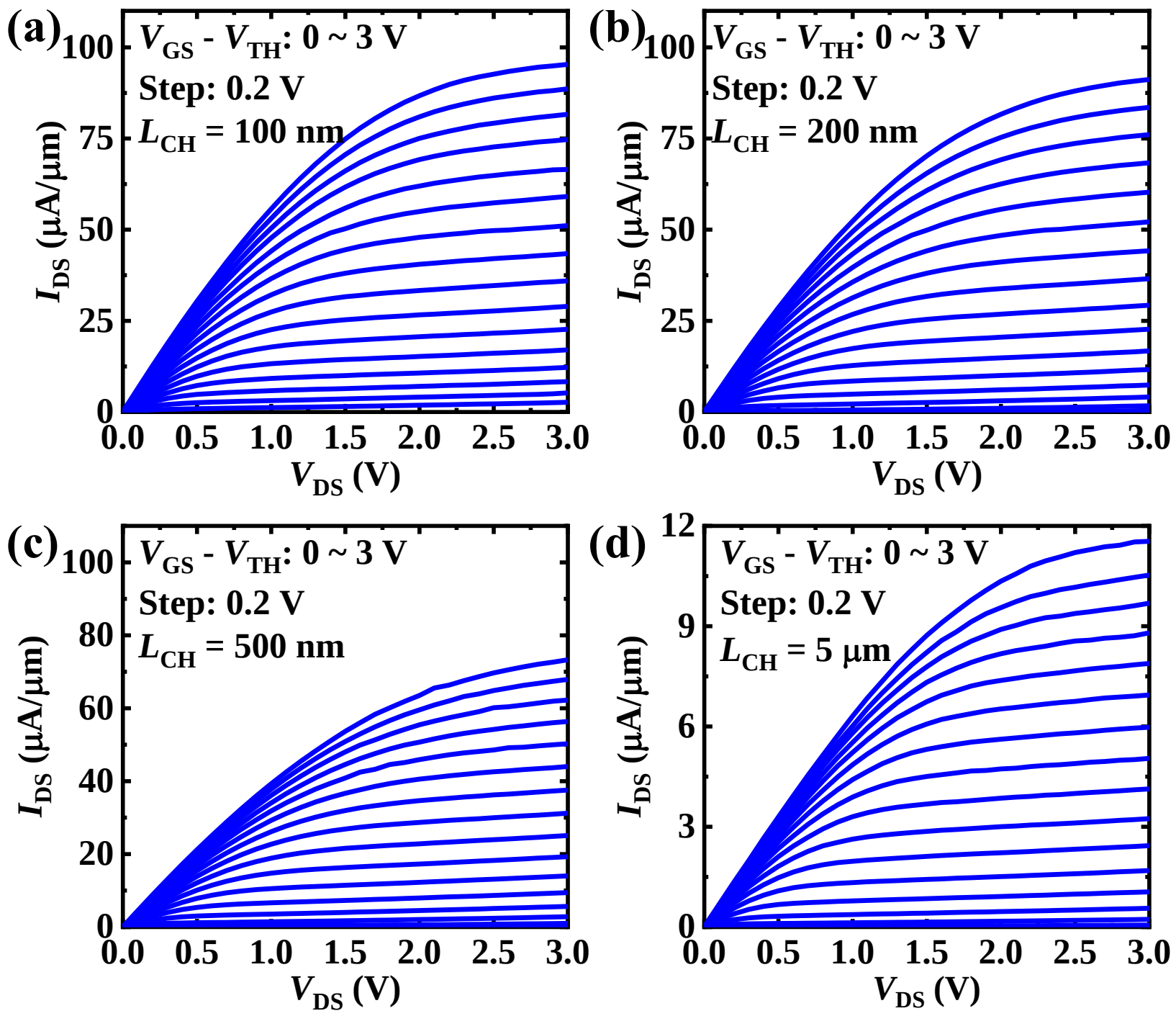

**Supplementary Fig. 6 Output characteristics of devices with different $L_{CH}$s.** The $I_{DS}$-$V_{DS}$ curves of devices with $L_{CH}$ of (**a**) 100 nm, (**b**) 200 nm, (**c**) 500 nm, and (**d**) 5000 nm are measured at the programmed state. $I_{ON}$ obtained at an overdrive voltage of 3 V and a $V_{DS}$ of 3 V saturates when $L_{CH}$ is below 500 nm.

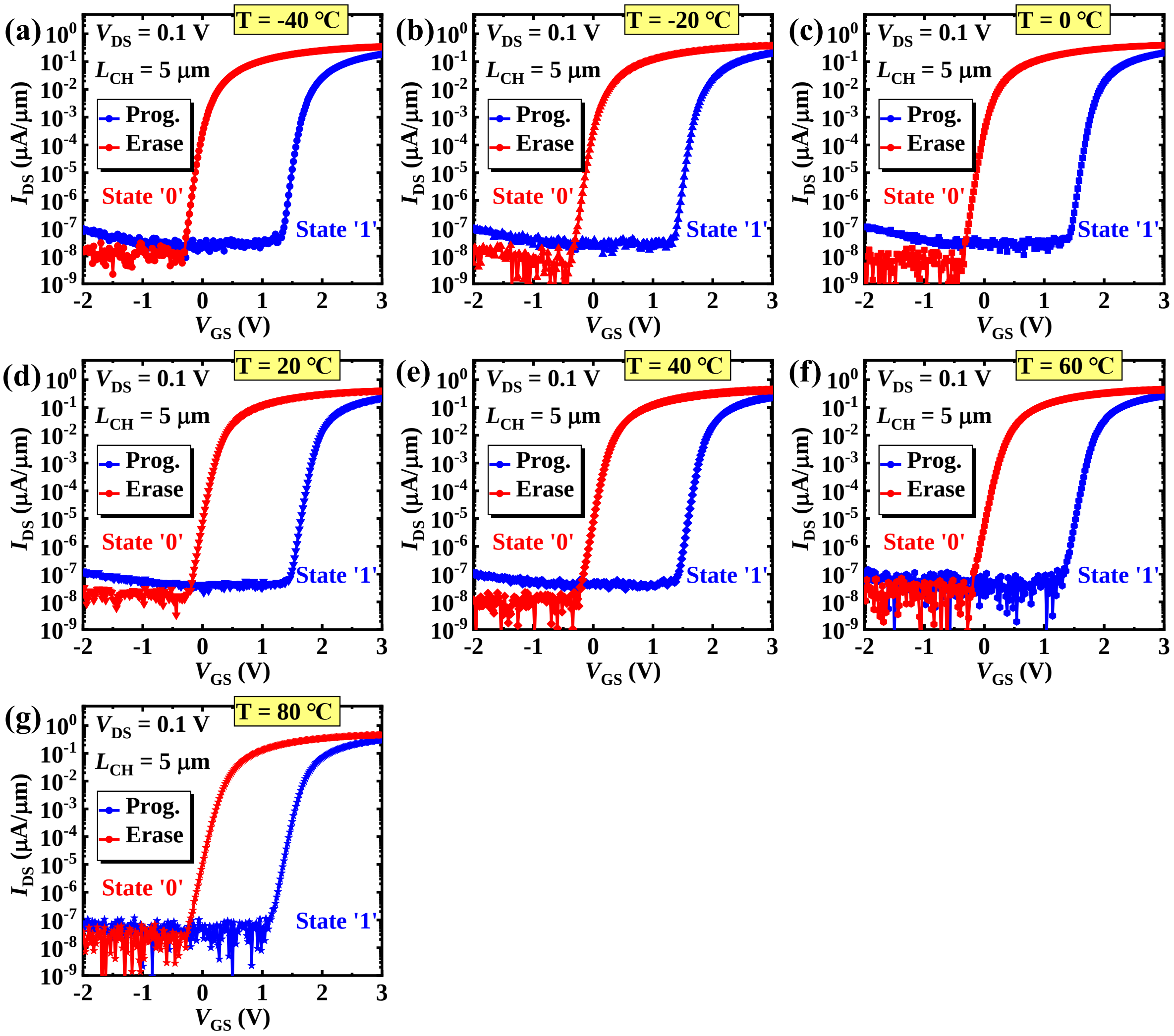

**Supplementary Fig. 7 Transfer characteristics of the FG *a*-IGZO TFT at different temperatures.** Device characteristics at temperature of (**a**) -40 °C, (**b**) -20 °C, (**c**) 0 °C, (**d**) 20 °C, (**e**) 40 °C, (**f**) 60 °C, and (**g**) 80 °C are measured under vacuum conditions.

| | $W_{CH}/L_{CH}$ (μm) | $I_{ON}$ (a) ($V_{OV}$,$V_{DS}$) (μA/μm) | $I_{ON}/I_{OFF}$ | Mobility ($cm^2$/V·s) | *SS* (mV/dec.) | MW (V) | Operate voltage (V) | Cycle | Retention (s) |
|---|---|---|---|---|---|---|---|---|---|
| **This work** | **1/0.06** | **127 (4,3)**<br>**7.2 (4,0.1)** | **>$10^8$** | **32.6** | **80-105** | **1.6–1.8** | **6** | >$10^3$ | >$10^8$ |
| 2021 Jeong[65] | 300/100 | 0.1 (6,0.3) | ~$10^5$ | **42** | 380 | 3.14 | 18 | >1000 | >$10^8$ |
| 2021 Bae[66] | **20/0.14** | 0.2 (20,0.1) | ~$10^2$ | - | ~750 | 10 | 20 | >3000 | >$10^4$ |
| 2021 Naqi[67] | - | 30 μA (b) (20,1) | ~$10^6$ | - | 2500 | 13.7 | 20 | >100 | >$10^7$ |
| 2020 Ryoo[68] | 100/50 | 0.02 (20,0.1) | ~$10^5$ | - | ~1200 | ~17 | 20 (c) | - | >$10^3$ |
| 2020 Kim[69] | **20/0.2** | 0.05 (15,0.1) | ~$10^6$ | 6.57 | ~1000 | 15 | 20 | >$10^4$ | >$10^4$ |
| 2020 Ma[70] | 10/10 | 1.1 (3,4) | **>$10^8$** | 10.96 | **112** | 8.2 | 20 | - | >$10^8$ |
| 2020 Liu[71] | 60/10 | 0.1 (6,0.1) | **~$10^8$** | - | 1000 | 8 | 13 | >$10^4$ | >$10^5$ |
| 2020 He[72] | 1000/80 | 0.01 (6,2) | ~$10^4$ | - | 500 | 7.3 | **6 (c)** | >10 | >600 |
| 2019 Yang[73] | 40/40 | 0.01 (12,1) | ~$10^5$ | - | 1110 | 20 | 20 (c) | - | >6000 |
| 2019 Yoon[74] | 40/40 | 0.03 (12,0.1) | **~$10^7$** | 6 | 200 | 25.6 | 20 | >8000 | >$10^4$ |
| 2019 Kim[75] | 40/20 | 0.075 (15,0.1) | ~$10^5$ | - | ~200 | 15.2 | 15 | >1000 | >$10^4$ |
| 2019 Son[76] | 40/20 | 0.05 (12,0.1) | **~$10^7$** | - | ~200 | 13.8 | 20 (c) | >1000 | >$10^3$ |
| 2019 Na[77] | 40/40 | 0.25 (15,0.5) | **~$10^7$** | - | ~200 | 25 | 20 (c) | - | >$10^4$ |
| 2018 Zhang[40] | 80/10 | 0.5 (3,5) | **$10^8$-$10^9$** | 6.1 | 198 | 5.74 | 16 | - | >$10^8$ |
| 2018 Yang[78] | - | 1 μA (b) (12,0.1) | ~$10^6$ | - | ~1000 | 19.8 | 20 (c) | - | - |
| 2018 Hwang[79] | 20/5 | 0.5 (8,5.1) | ~$10^6$ | 19.3 | 182 | ~2 | 20 | >$10^4$ | >$10^8$ |
| 2018 Koo[80] | 1000/100 | 0.01 (2,1.5) | ~$10^4$ | - | ~300 | 0.88 | **9** | >30 | >$10^4$ |
| 2017 Ji[38] | 30/20 | 0.3 (8,-) | ~$10^5$ | - | ~1300 | 3.7 | 13 | >400 | >$10^8$ |
| 2017 Seo[81] | 40/40 | 0.05 (15,0.1) | ~$10^6$ | - | **150** | 18 | 17 | >7000 | >$10^4$ |
| 2017 Qian[82] | 100/10 | ~0.1 (4,0.1) | $10^6$ | - | ~400 | 8.03 | 18 | - | >$10^5$ |
| 2017 Qian[83] | 100/10 | 0.04 (6,0.1) | ~$10^6$ | 7.1 | 700 | 4.7 | 18 | - | >$10^8$ |
| 2017 Park[84] | - | ~3 μA (b) (12,0.1) | **~$10^7$** | - | ~500 | 8 | 20 (c) | - | - |
| 2016 Ahn[85] | 10-Oct | 0.8 (15,1) | $10^5$-$10^6$ | - | ~2000 | 11.1 | 14 | >$10^4$ | >$10^8$ |
| 2016 Yun[86] | 40/40 | ~0.05 (12,0.1) | **~$10^7$** | 8.71 | 230 | 6.2 | 20 | - | >$10^4$ |
| 2016 Hanh[87] | 80/40 | 0.05 (6,1) | ~$10^6$ | 16.3 | 1150 | 2.12 | 11 | - | >$10^8$ |
| 2016 Pan[88,89] | 100/10 | 3 (8,5) | **~$10^7$** | - | 196 | 4.1 | 20 | >$10^5$ | >$10^8$ |
| 2016 Kim[90] | 40/20 | 2.5 (10,5.5) | **~$10^7$** | 6 | 250 | 25.6 | 25 (c) | >8000 | >$10^4$ |
| 2015 Qian[39] | 100/10 | 0.7 (4,1) | **~$10^8$** | 14.6 | 440 | 4.67 | 19 | >1000 | >$10^5$ |
| 2015 Kim[91] | 40/20 | 0.1 (15,0.1) | **~$10^8$** | - | 320 | 25.8 | 30 (c) | - | >$10^4$ |
| 2015 Zhang[92] | 50/10 | 0.1 (2,1) | ~$10^6$ | 2 | 840 | 4.68 | 15 | > 100 | >$10^5$ |
| 2015 Bak[93] | 40/20 | 0.05 (15,0.1) | **~$10^7$** | 0.13 | ~200 | 18.4 | 20 (c) | >$10^4$ | >$10^5$ |
| 2015 Li[94] | 96/24 | 0.01 (12,0.1) | ~$10^5$ | 6.2 | 240 | 14.4 | 20 | - | >$10^8$ |
| 2015 Her[95] | 100/10 | 0.4 (6,1) | ~$10^6$ | - | ~600 | 2.7 | 15 | >$10^4$ | >$10^5$ |
| 2014 Bak[96] | 40/20 | 0.15 (8,1) | **~$10^8$** | 0.51 | 410 | 7.7 | 20 | - | >$10^8$ |
| 2014 Bak[97,98] | 40/20 | 0.05 (20,0.1) | **~$10^7$** | - | ~400 | 17.1 | 20 | >$10^4$ | >$10^4$ |
| 2014 Chen[99] | 100/35 | 0.002 (3,0.1) | ~$10^4$ | 2.28 | 665 | 1.99 | 12 | - | >$10^4$ |
| 2013 Cui[100] | 50/5 | 0.07 (2,5) | ~$10^6$ | 8.4 | 390 | 6.15 | **10 (c)** | >100 | >$10^8$ |
| 2013 Chen[101] | 100/25 | 0.0025 (2.5,8) | ~$10^4$ | - | ~600 | 6.38 | 15 | - | >$10^4$ |
| 2012 Nguyen[102] | 80/12 | 0.05 (5,1) | **~$10^7$** | - | ~200 | 3 | **10** | - | >$10^8$ |
| 2012 Jung[103] | 20/5 | - | - | - | - | ~8 | 20 | - | >$10^5$ |
| 2011 Jang[104] | 50/10 | 0.1 (15,0.1) | **~$10^7$** | - | 1200 | 4.7 | 35 | >1050 | |
| 2011 Nguyen[105] | 64/32 | 0.1 (3,1)) | **~$10^7$** | 4.18 | **130** | 3.19 | 14 | - | - |
| 2010 Park[106] | 50/5 | 0.0003 (30,20) | ~$10^3$ | - | 8000 | 15 | 50 | >200 | >$10^8$ |
| 2010 Su[107] | 500/50 | 0.015 | ~$10^6$ | - | **128** | 1.6 | 12 | >5000 | >$10^7$ |
| 2009 Suresh[108] | 400/100 | 0.25 (10,10) | **~$10^7$** | 14 | 200 | 5.1 | **10 (c)** | - | >$10^8$ |
| 2008 Yin[109] | 50/4 | 0.5 (4,1.1) | **~$10^8$** | 10.3 | 220 | 3.5 | 11 | - | >$10^4$ |
| 2008 Yin[110] | 50/4 | 0.52 (4,1.1) | **~$10^8$** | 10.3 | 207 | 3.8 | 12 | >1000 | >$10^4$ |

a: Overdrive voltage: $V_{OV} = V_{GS} - V_{TH}$
b: Not normalized to the channel width.
c: Obtained by DC sweep, not by applying pulses

**Supplementary Table 1 Performance benchmark table for flash memories using the *a*-IGZO channel.** Devices reported in this work show the shortest $L_{CH}$ with the best control of short-channel effects ($SS$ = ~105 mV/dec. for $L_{CH}$ = 60 nm), achieving the highest $I_{ON}$ of 127 $\mu$A/$\mu$m ($V_{OV}$ = 4 V, $V_{DS}$ = 3 V) and the lowest operating voltage of 6 V among all the flash memories based on an *a*-IGZO channel. The high $I_{ON}$ and ultra-low $I_{OFF}$ also result in the high $I_{ON}/I_{OFF}$ of more than 8 orders.

| **TCAM state** | | **Search bit (SL, $\overline{SL}$)** | **ML** | **Condition** |
|---|---|---|---|---|
| **Logic** | **($D_0$, $D_1$)** | | | |
| **0** | **(0, 1)** | **0 (0, 1)** | **1** | **Match** |
| | | **1 (1, 0)** | **0** | **Mismatch** |
| **1** | **(1, 0)** | **0 (0, 1)** | **0** | **Mismatch** |
| | | **1 (1, 0)** | **1** | **Match** |
| **X** | **(1, 1)** | **0 or 1** | **1** | **Match** |

**ML = XNOR (TCAM state, Search bit)**

**Supplementary Table 2 Operation schemes of conventional 16T-CMOS TCAM cell.** The logic state (0, 1, X) of the TCAM cell is indicated by the states of SRAM. The logic state of the search bit is indicated by the states of search lines (SL and $\overline{SL}$). The output of the ML is determined by the XNOR result of the stored TCAM state and the search bit.

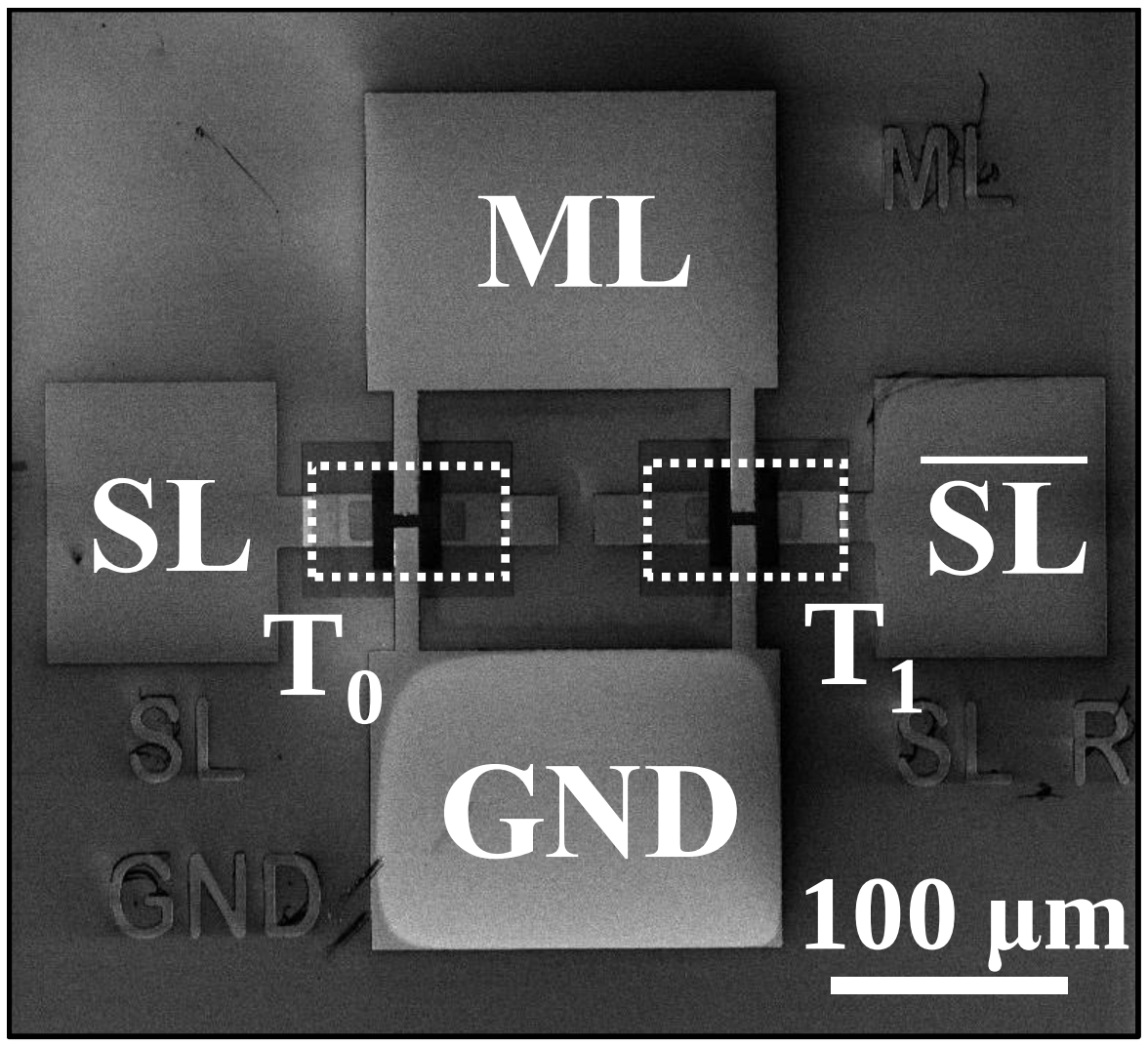

**Supplementary Fig. 8 Top-view SEM image of the TCAM cell.** The TCAM cell consists of two parallel-connected FG *a*-IGZO TFTs.

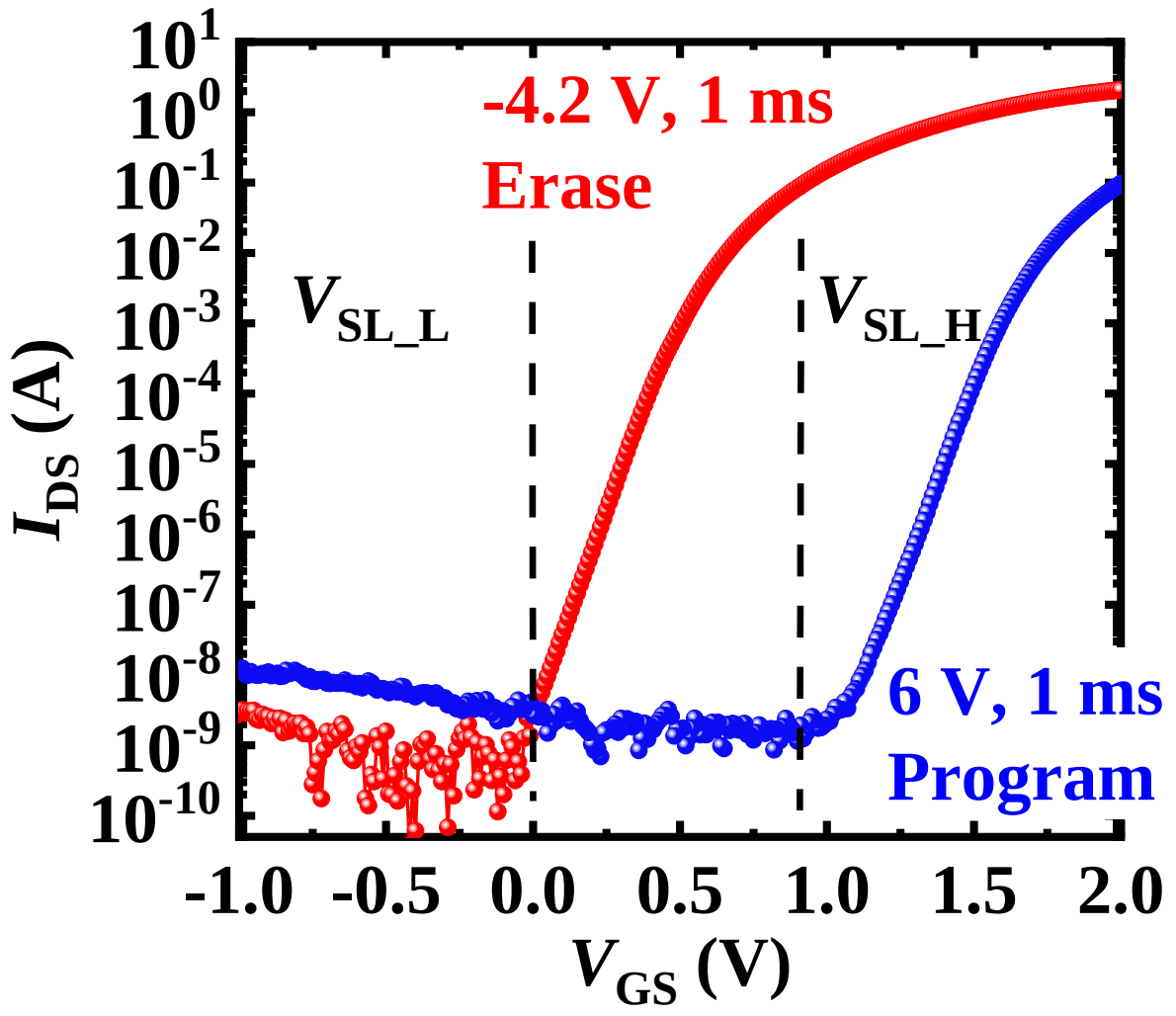

**Supplementary Fig. 9 $I_{DS}$-$V_{GS}$ curves after applying pulses during TCAM measurement.** The negative erase pulse (-4.2 V, 1 ms) only partially shifts the $V_{TH}$ of the transistor. If the transistor is entirely erased, it can be switched on even at the search voltage of 0 V (Another method is to apply a lower $V_{SL_L}$).

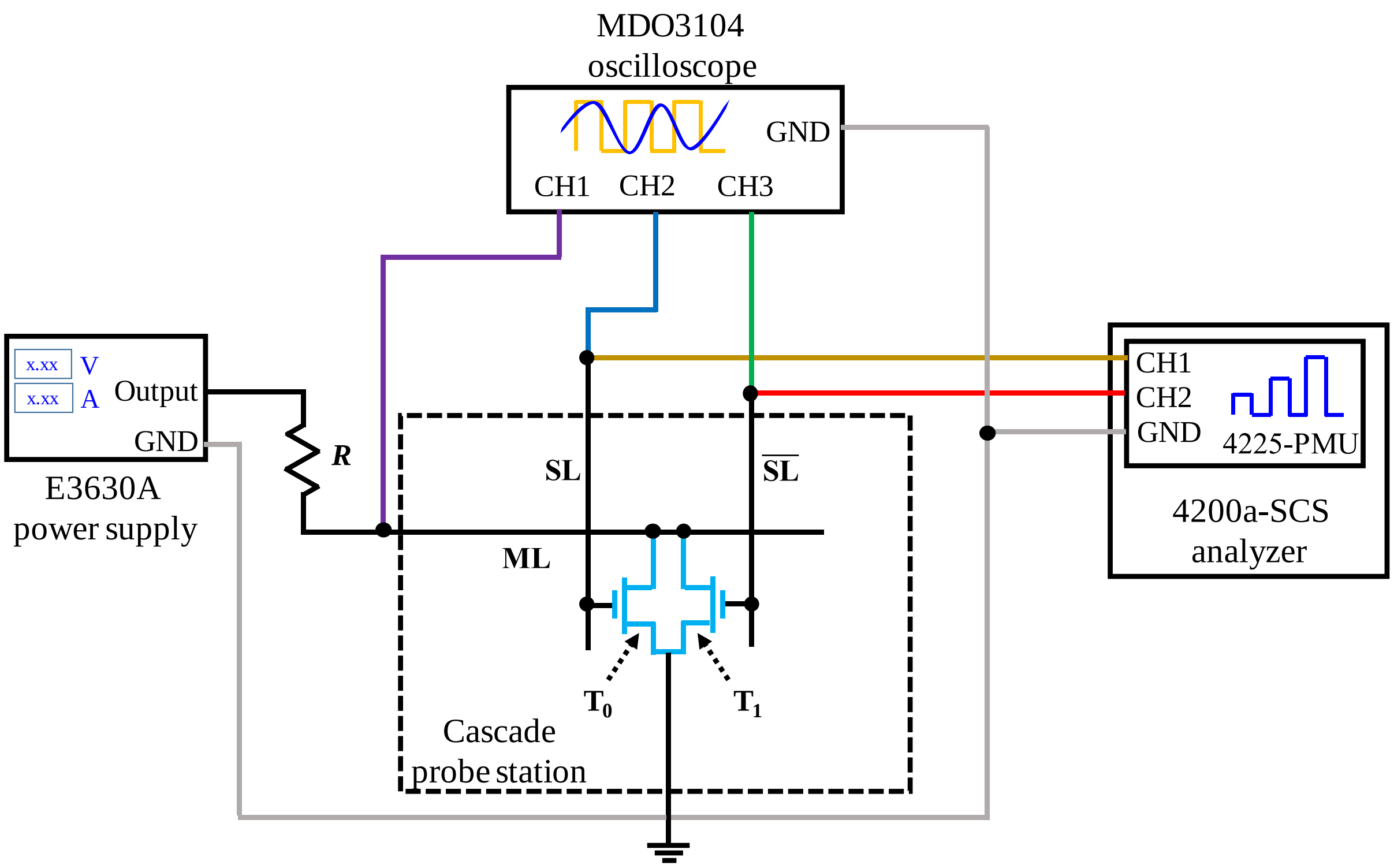

**Supplementary Fig. 10 Schematic of the measurement setup of TCAM cell.** The ML of the TCAM cell is connected to an external resistor (3 MΩ). The other terminal of the resistor is connected to the power supply with a $V_{DD}$ of 1.2 V. Input pulses for writing and searching are generated by the two-channel pulse measurement unit (PMU). The waveforms of the search lines and the ML are plotted by the oscilloscope.

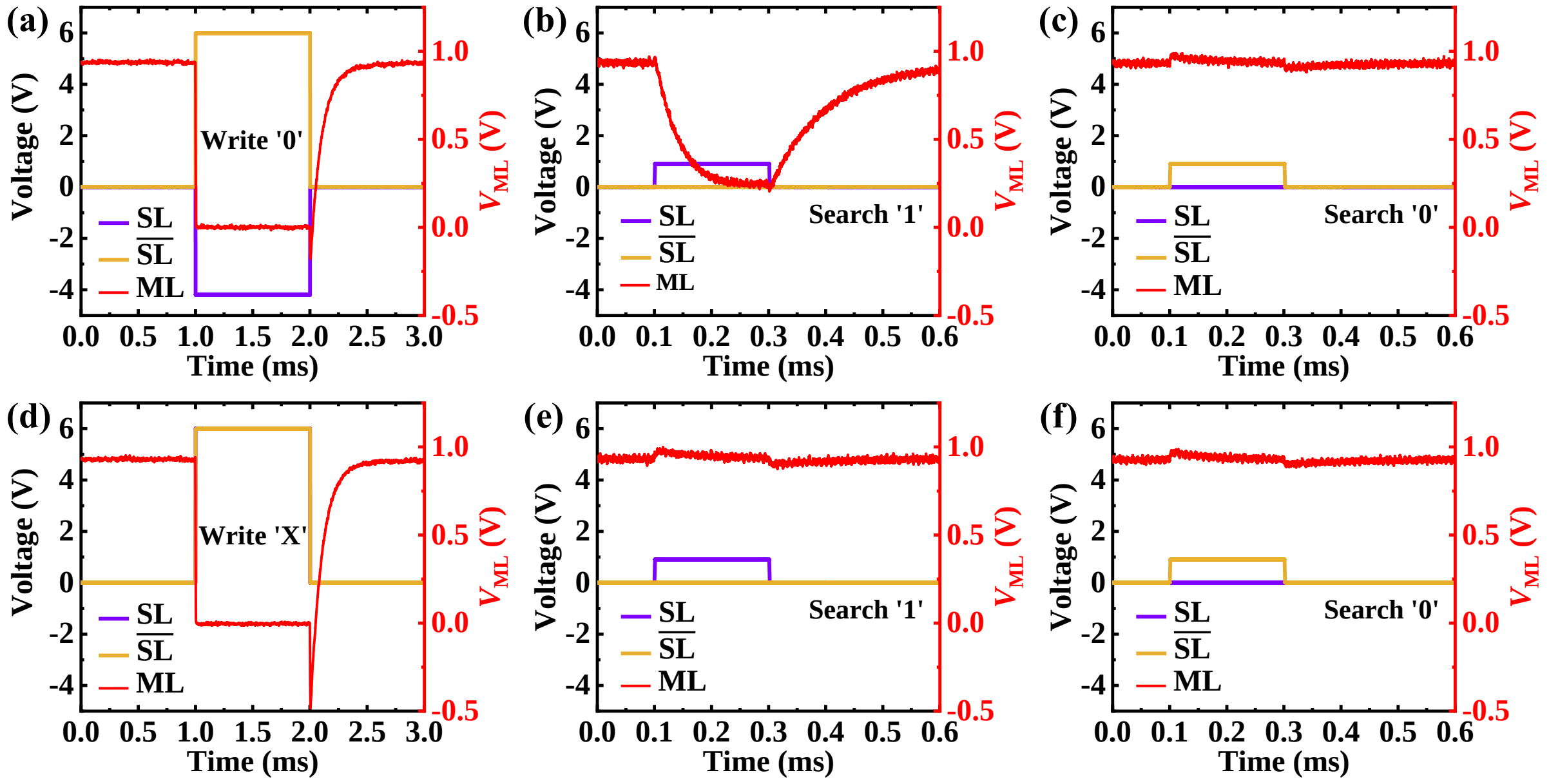

**Supplementary Fig. 11 Writing '0' and 'X' and corresponding search schemes. a**. Operation of writing '0' and corresponding search (**b**) '1', and (**c**) '0' schemes. **d**. Operation of writing 'X' and corresponding search (**e**) '1', and (**f**) '0' schemes. The ML keeps high for both searching '1' and '0' after writing 'X'.

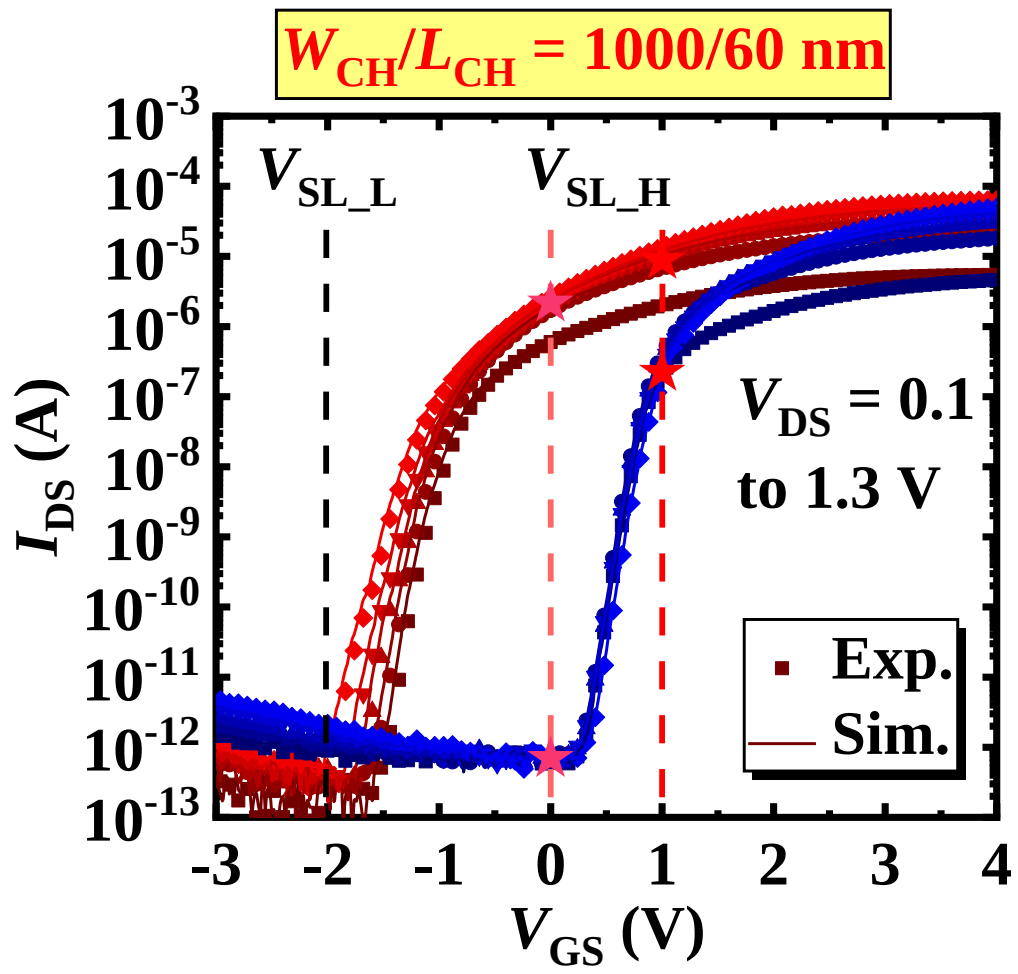

**Supplementary Fig. 12 Modeling and fitting of FG *a*-IGZO TFT ($L_{CH}$ = 60 nm) for array-level simulation.** The model used for simulations is well-calibrated according to the experimental data. $V_{SL_L}$ is set to -2 V, while $V_{SL_H}$ is set to vary from 0 V to 1 V.

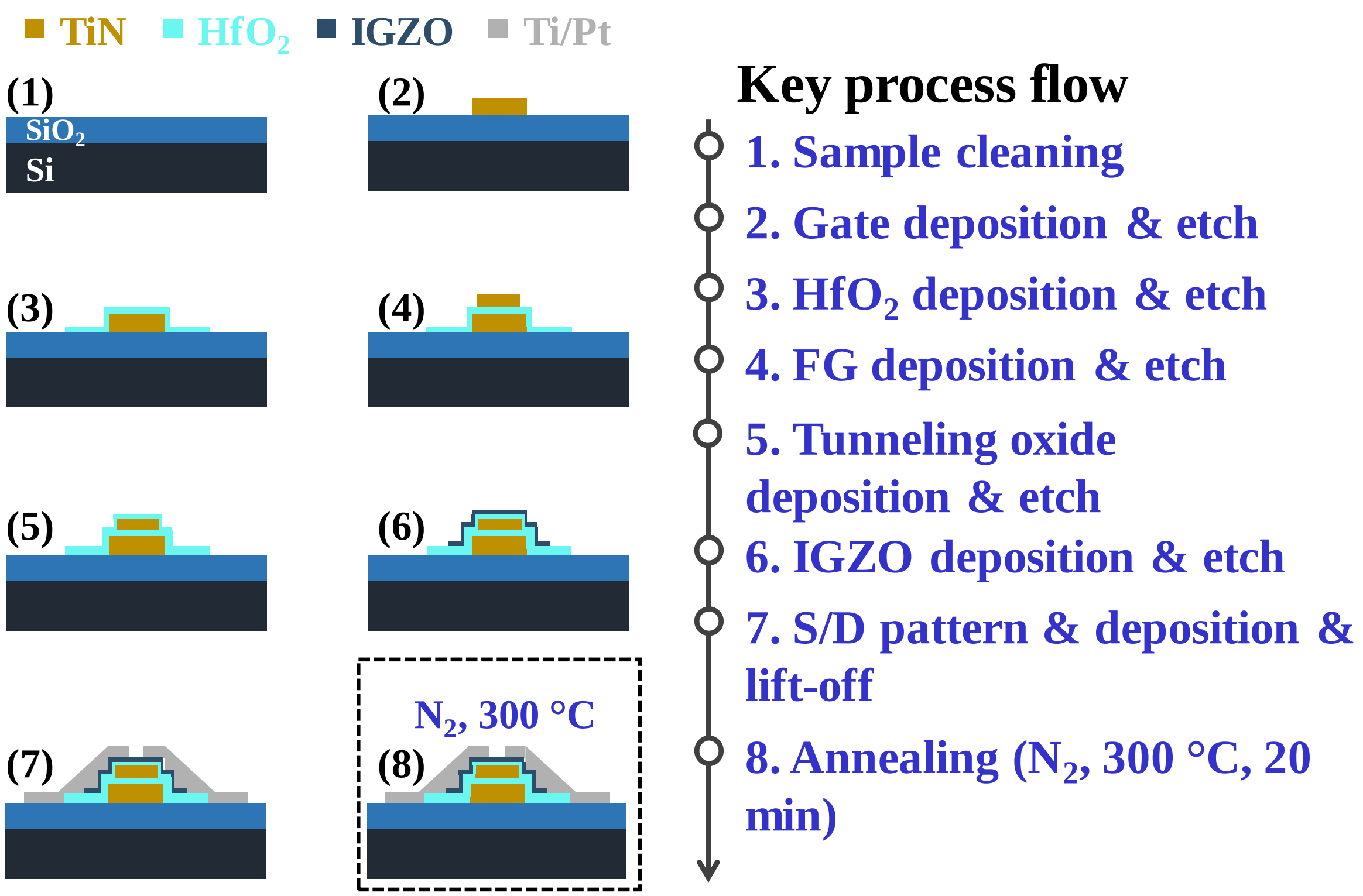

**Supplementary Fig. 13 Key process steps of the FG *a*-IGZO TFT and TCAM.** The TCAM cell can be formed during source/drain patterning by connecting the source and drain terminals of two TFTs, respectively.